\documentclass[letterpaper,10pt]{extarticle}
\usepackage{url} 
\usepackage{soul}
\usepackage{mathtools}
\usepackage{setspace}
\usepackage{placeins}

\usepackage{natbib}
\setcitestyle{authoryear,open={(},close={)}}
\usepackage{local}
\usepackage[left=1in,top=.65in,bottom=.65in,right=1in]{geometry}
\usepackage{hyperref}
\usepackage{xcolor}

\title{Spatial correlation  increase in single-sensor satellite data reveals loss of Amazon rainforest resilience}
\author{Lana L. Blaschke,\textcolor{Accent}{\textsuperscript{1,2*}} 
	Da Nian,\textcolor{Accent}{\textsuperscript{2}} 
	Sebastian Bathiany,\textcolor{Accent}{\textsuperscript{1,2}}  
	Maya Ben-Yami,\textcolor{Accent}{\textsuperscript{1,2}} 
	Taylor Smith,\textcolor{Accent}{\textsuperscript{3}} 
	Chris A. Boulton,\textcolor{Accent}{\textsuperscript{4}} 
	Niklas Boers,\textcolor{Accent}{\textsuperscript{1,2,4,5}}  \\
	\begin{small}
		\textcolor{Accent}{\textsuperscript{1}}Earth System Modelling, School of Engineering and Design, Technical University of Munich, Munich, Germany\\ 
		\textcolor{Accent}{\textsuperscript{2}}Potsdam Institute for Climate Impact Research, Potsdam, Germany\\
		\textcolor{Accent}{\textsuperscript{3}}Institute of Geosciences, University of Potsdam, 14476 Potsdam, Germany\\
		\textcolor{Accent}{\textsuperscript{4}}Global Systems Institute, University of Exeter, EX4 4QE Exeter, UK\\
		\textcolor{Accent}{\textsuperscript{5}}Department of Mathematics, University of Exeter, EX4 4QF Exeter, UK\\
		\textcolor{Accent}{\textsuperscript{*}}Correspondence: \textcolor{Accent}{lana.blaschke@pik-potsdam.de} \\ 
	\end{small}
}	

\begin{document}
	
\maketitle
\bigskip

\begin{abstract}
	\doublespacing
	The Amazon rainforest (ARF) is threatened by deforestation and climate change, which could trigger a regime shift to a savanna-like state. Previous work suggesting declining resilience in recent decades was based only on local resilience indicators. Moreover, previous results are potentially biased by the employed multi-sensor and optical satellite data and undetected anthropogenic land-use change.
	Here, we show that the spatial correlation provides a more robust resilience indicator than local estimators and employ it to measure resilience changes in the ARF, based on single-sensor Vegetation Optical Depth data under conservative exclusion of human activity. Our results show an overall loss of resilience until around 2019, which is especially pronounced in the southwestern and northern Amazon for the time period from 2002 to 2011. The demonstrated reliability of spatial correlation in coupled systems suggests that in particular the southwest of the ARF has experienced pronounced resilience loss over the last two decades.
\end{abstract}

\pagebreak

\section*{Introduction}

\begin{figure}
	\centering
	\includegraphics[width=.6\textwidth]{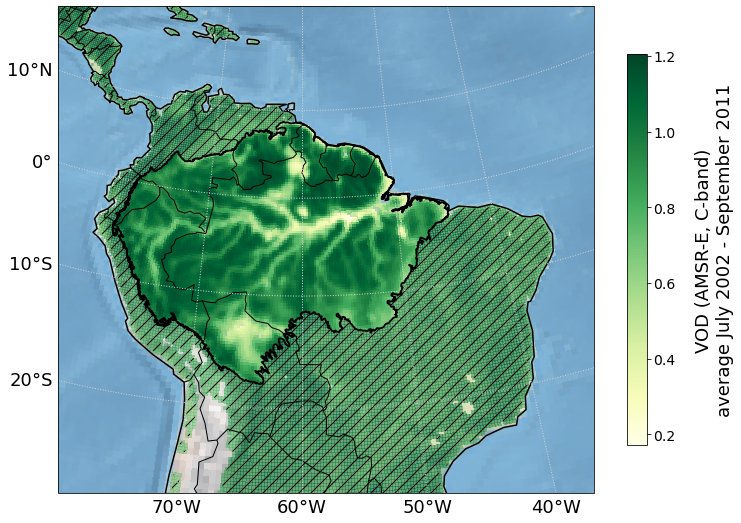}
	\caption{
		\textbf{Average Vegetation Optical Depth in the Amazon basin measured by the C-band of the AMSR-E satellite sensor.}
		Daily data is aggregated to a monthly resolution by taking the mean over complete months, hence the time period from July 2002 to September 2011 is considered in the case of AMSR-E.
		The outline of the Amazon basin can be found at 
	}
	\label{fig:AMSRE_mean_years}
\end{figure}

The Amazon rainforest (ARF) is the most biodiverse region of our planet, and serves as a major carbon sink \citep{malhi_climate_2008,gatti_amazonia_2021}. Yet, the efficiency of its carbon uptake has been declining over the last decades \citep{malhi_climate_2008,gatti_amazonia_2021,brienen_long-term_2015,hubau_asynchronous_2020}, with the ARF becoming carbon neutral and even acting as a carbon source during the two one-in-a-century droughts in 2005 and 2010  \citep{phillips_drought_2009,gatti_drought_2014,feldpausch_amazon_2016}.
The ARF's important role in the global carbon cycle thus means that its existence and stability are crucial for climate change mitigation, especially as the planet continues to warm in the later parts of the century \citep{shuka_climate_2022}.

Studies suggest that there is a critical mean annual precipitation (MAP) value at which parts of the forest might irreversibly transition into a savanna-like state \citep{hirota_global_2011,zemp_self-amplified_2017}. 
In such a scenario, forest dieback would likely be self-amplifying, i.e. the non-linearity of such an abrupt transition would result from positive feedback mechanisms operating in the region.
Besides fire \citep{brando_abrupt_2014}, the main feedback mechanism that could amplify dieback in the ARF is related to moisture recycling  \citep{staal_forest-rainfall_2018,salati_recycling_1979}. 
Moisture is transported at low atmospheric levels via the trade winds from the tropical Atlantic to the Amazon basin, where it precipitates. A substantial fraction is taken up by the vegetation and transpired back to the atmosphere, or evaporates from the complex surfaces of plants. This evapo-transpirated water is then transported further west and south over the Amazon and towards the Andes by low-level jets, sometimes called atmospheric rivers \citep{gimeno_major_2016}. The low-level circulation itself is amplified by condensational latent heating over the Amazon basin, strengthening the large-scale atmospheric heating gradient between ocean and land \citep{boers_deforestation-induced_2017}.

Two main mechanisms have been proposed that may activate positive feedback cycles and push the ARF towards a critical threshold \citep{lovejoy_amazon_2018,lovejoy_amazon_2019}.
On the one hand, anthropogenic global warming will cause increased temperatures over the Amazon basin, which could lead to increased evapo-transpirative demand without a corresponding increase in water supply via precipitation, especially during a potentially intensifying and prolonging dry season \citep{malhi_exploring_2009} and severe droughts \citep{vogel_projected_2020}. This could additionally lead to decreased convection and a reduction of moisture inflow from the Atlantic \citep{pascale_current_2019}; moreover, models from the Coupled Model Intercomparison Project Phase 6 (CMIP6) project an overall drying in tropical South America in response to increasing atmospheric greenhouse gas concentrations. Hence, global warming could drive the system towards destabilization \citep{zemp_self-amplified_2017}.
Furthermore, deforestation can lead to a critical decrease of evaporated moisture transported downstream, and to an additional reduction of moisture inflow due to a decreased heating gradient, further pushing the ARF toward a critical threshold \citep{boers_deforestation-induced_2017,zemp_deforestation_2017}. The decrease in precipitation that would occur beyond such a threshold would also cause degradation outside the ARF region. 

Such vegetation carbon losses have been predicted by several CMIP6 models in parts of the Amazon basin, preceded by an increasing amplitude of seasonal temperatures \citep{parry_evidence_2022}.
Furthermore, observations have shown that the Amazonian dry season is increasing in length \citep{marengo_meteorological_2017,leite-filho_effects_2019,phillips_drought_2009}, exacerbated by the three severe droughts that have occurred since 2005 \citep{feldpausch_amazon_2016}.
In view of these projected and observed trends and the global relevance of the ARF, monitoring changes in its resilience is of great importance.

As the data-driven monitoring of resilience changes and the anticipation of critical transitions is important for many parts of our climate, including the ARF, a considerable amount of research has focused on developing and applying such methods. 
Existing methods focus on the detection of distinct signs of resilience loss, where resilience is defined as a system's ability to recover from perturbations; the underlying mathematical concept is derived from dynamical system theory.
Under the assumption that resilience loss can be dynamically represented by an approaching (codimension-1) bifurcation, the approach to the critical forcing value at which the bifurcation is accompanied by a weakening of the equilibrium restoring forces in the system, resulting in slower recovery from small perturbations. This is termed `critical slowing down' (CSD). During CSD, the variance and lag-1 autocorrelation (AC1) of the system state increase, as they are directly linked to the recovery rate from perturbations \citep{scheffer_early-warning_2009,dakos_slowing_2008,dakos_robustness_2012,boers_theoretical_2022}. 
Yet, in the case of spatio-temporal data, calculating the variance and AC1 of individual point locations does not exploit the information that is potentially encoded in the spatial dimensions, and in particular misses the interactions between different locations.
In \citet{dakos_spatial_2010} the authors argue that when a system consists of several coupled units, a decrease in the units' recovery rates causes increasing correlation between two coupled cells.

Variance and AC1 have recently been confirmed to quantify resilience empirically at global scale, by comparing theoretical estimates of the recovery rates based on classical CSD indicators to direct estimates of the recovery rate from VOD time series sections showing recovery from abrupt disturbances \citep{smith_empirical_2022}. 
Recent results in \citet{boulton_pronounced_2022} have revealed that large parts of the ARF's vegetation biomass show increasing AC1, implying a loss of resilience that has been especially pronounced since the early 2000s.
The studies in \citet{smith_empirical_2022} as well as \citet{boulton_pronounced_2022} used multi-satellite data to extend the period of time under study \citep{moesinger_global_2020}.
Yet, \citet{smith_reliability_2023} subsequently showed that such merging of sensors can create spurious changes in higher-order statistics such as variance and AC1 when those are calculated from multi-instrument time series.
The non-stationarity induced in the time series by the change of sensors with different orbits, intrinsic noise, and radiometric resolutions may result in statistical signals which can be misinterpreted as resilience changes. It is thus recommended to investigate the resilience changes in a system by using single-sensor instrument records when available.
In this work we thus exclusively use single-sensor data: in particular, Amazon vegetation indices based on Vegetation Optical depth (VOD), see Fig. \ref{fig:AMSRE_mean_years}.
While indicators of CSD could react differently to changes in the measurement process or to systemic changes that are not related to CSD, different indicators that are related to CSD via the recovery rate should behave consistently whenever resilience changes. We thus ensure the robustness of our results by comparing the calculated spatial correlation to corresponding estimates of the AC1 and variance.
To further ensure robustness, we compare different data sources as recommended in \citet{samanta_why_2012}.

It has been suggested that different areas of the ARF can be considered part of a coupled system connected by spatial interactions via evapo-transpiration, low-level winds, and moisture recycling \citep{staal_forest-rainfall_2018}. Since the trade winds transport moisture from east to west, we assume that coupling cells uni-directionally is a reasonable simplification of the plant-water moisture transport system in the Amazon \citep{boers_deforestation-induced_2017}. The almost laminar flows can be investigated as separate trajectories, thereby allowing a reduction to one average dimension.
Based on this relationship, we set up a simple conceptual model with asymmetric interaction (see Fig 2.) to validate  variance, AC1, and spatial correlation as indicators of CSD and, hence, resilience loss in ARF vegetation.
We then calculate all three resilience indicators for four single-sensor VOD satellite data sets, which quantify the ARF's vegetation proxy of above-ground water content in biomass. 
Finally, based on these results we discuss changes in resilience of the ARF.

\section*{Materials and Methods}
\subsection*{Description and parametrization of the conceptual model}
\label{sec:methods:VECODE}

\begin{figure}
	\centering
	\includegraphics[width=\textwidth]{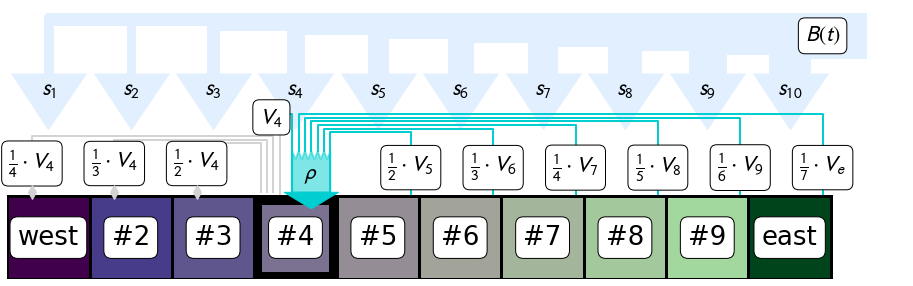}
	
	\caption{
		\textbf{Conceptual moisture recycling model.} 
		The thin turquoise arrows represent the incoming precipitation due to moisture recycling and the thick light blue arrows indicate the precipitation directly originating from the Atlantic ocean.
		For each cell $i$, a scaling factor $s_i$ determines the fraction of `background precipitation' $B(t)$ that precipitates.
		As an example, cell $\#4$ and its corresponding incoming and outgoing fluxes are highlighted.
		Each cell to the east of cell $\# 4$ contributes by ${(d_{i}+1)}^{-1} \cdot V_i$ where $d_i$ is the distance between cell $i$ and cell $\# 4$ and $V_i$ is the moisture content in cell $i$.
		Thin grey arrows mark the precipitation in other cells that originated from cell $\# 4$.
		The decrease of the control parameter $B(t)$ representing the overall incoming moisture from the Atlantic induces a critical transition in the vegetation model. 
	}
	\label{fig:VECODE_scheme}
	
\end{figure}

To demonstrate the relevance of the three indicators of CSD in a setting like the ARF, 
we slightly modify a previously introduced conceptual model of atmosphere-vegetation interaction 
\citep{bathiany_detecting_2013-1}.
The vegetation dynamics of the model are inspired by the global dynamical vegetation model VECODE
\citep{brovkin_continuous_1997, brovkin_carbon_2002,brovkin_stability_1998,bathiany_detecting_2013}, which is based on an empirical relationship of vegetation cover fraction and atmospheric conditions. 

In particular, the equilibrium vegetation $V^*$ is a monotonic (sigmoidal) function of precipitation, see Equation~\ref{eq:V*}.
The vegetation dynamics are simulated as a linear relaxation toward this empirical equilibrium and the equilibrium vegetation $V^*$ is a direct function of total incoming precipitation $P_i$ in [mm/year], namely
\begin{linenomath*}
\begin{equation}
	\label{eq:V*}
	V^* =
	\begin{cases}
		0 & \text{if } P_i < P1\\
		1 & \text{if } P_i > P2\\
		1.03 - 1.03 \left( 1+ \frac{\alpha}{\text{exp}(\phi)^2} (P_i - P1)^2\right)^{-1}
		& \text{otherwise.}
	\end{cases}       
\end{equation}
\end{linenomath*}
Here, $P_i$ represents the amount of precipitation in a cell $i$,  as the spatial extension of our model consists of 10 cells. 
While VECODE is based on an empirical relationship of vegetation cover fraction based on atmospheric conditions, we here interpret the vegetation variable $V^*$ as a property linked to the tree cover in the Amazon (e.g., tree cover fraction or biomass). We motivate this interpretation by the conceptual nature of the model approach, and the fact that tree coverage in the original VECODE version is modeled by a very similar approach as vegetation coverage (with a curve that is shifted to higher precipitation rates). Making the model more complex by distinguishing plant types would hence not add to our analysis.

The parameters $P1$ and $P2$ in Eq. \ref{eq:V*} are given by
\begin{linenomath*}
\begin{equation}
	\begin{aligned}
		P1 &= \beta \cdot \text{exp}(\frac{\phi}{2}) \\
		P2 &= \beta \cdot \text{exp}(\frac{\phi}{2}) + \frac{\text{exp}(\phi)}{\sqrt{0.03 \alpha}}
		.
	\end{aligned}
\end{equation}
\end{linenomath*}
Parametrization is chosen in a similar way to \citet{bathiany_detecting_2013-1} but adapted such that the range of bistability in dependence of precipitation is closer to that in the ARF. Namely, the parameters are $\alpha = 0.0011$, $\beta = 280$, and $\phi = 2.45$. 
Thus, the vegetation $V_i \in [0,1]$  of a cell $i$ at time $t+1$ is given by
\begin{linenomath*}
\begin{equation}
	\label{eq:dV}
	V_i^{t+1} = V_i^{t} +  \frac{V^*(P_i^{t}) - V_i^{t}}{\tau} \Delta t 
	,
\end{equation}
\end{linenomath*}
where $P_i^{t}$ is itself a function of the vegetation state $V^{t}$. Namely, following \citet{bathiany_detecting_2013-1}, we implement an atmosphere-vegetation feedback in our model by forcing the precipitation with the vegetation, installing the moisture-advection feedback as the coupling mechanism.
While this coupling can in principle capture many land surface mechanisms, with this focus, the resulting feedback mechanisms can then be considered to act uni-directional from east to west following the fluxes in the atmosphere over the Amazon \citep{staal_forest-rainfall_2018, salati_recycling_1979}.
Thus, the cells can be thought as in a row with cell \#1 representing the most downstream region (southwestern Amazon) and cell \#10 the eastern-most cell (East coast of South America).
Then, the precipitation from moisture recycling $P_i^{recycled}$ is a sum of vegetation in the cells to the east of cell $i$ (cells with higher indices $j\geq i$), weighted by the distances. Mathematically, it can be expressed as
\begin{linenomath*}
\begin{equation}
	\label{eq:P_i_recycled}
	P_i^{\text{recycled}} = \rho \cdot \sum_{j\geq i} \frac{1}{j-i+1} \cdot V_j
\end{equation}
\end{linenomath*}
and acts a coupling between the cells. The scaling factor $\rho$ is set to $600$.
Note that each cell's precipitation is also dependent on its own vegetation cover $V_i$.
Additionally, in each cell the precipitation depends on the amount of moisture that arrives from the Atlantic ocean, which is represented by the control parameter $B$, as well as on the amount of its own and more eastern cells' vegetation.
This background precipitation is the product of a scaling factor $s_i$ times $B(t)$. The scaling factor 
\begin{linenomath*}
\begin{equation*}
	s_i = max\{0.1\cdot i, \ 0.2\}
\end{equation*}
\end{linenomath*}
resembles the amount of precipitation that results directly from the ocean. Hence, it is highest in the east ($s_{10} = 1.0$) and decreases towards the west ($s_1=s_2=0.2$). 
In summary, the total precipitation in cell $i$ is a sum of background precipitation and precipitation $P_i^{recycled}$ from moisture recycling in the `east' of the cell and given by 
\begin{linenomath*}
\begin{equation}
	\label{eq:P_i}
	P_i = B \cdot s_i + P_i^{\text{recycled}} + \sigma \eta ,
\end{equation} 
\end{linenomath*}
where the standard deviation of the white noise is set to $\sigma = 20$.
A visualization of the deterministic part of precipitation in the model is given in Fig. \ref{fig:VECODE_scheme}.

To mimic a climate change scenario with declining moisture inflow from the Atlantic ocean, the bifurcation parameter $B$ decreases linearly from 1100 to 990 over time in our model runs. 
One should note that due to the different magnitudes of moisture inflow from moisture recycling, the critical value of the control parameter $B$ differs between the cells. 
The interaction of vegetation and precipitation leads to the stability diagram shown in Fig. \ref{fig:SI_VECODE_stability} and to a transition to a low vegetation state once the critical threshold of precipitation is crossed.


While the characteristic relaxation time in VECODE is climate dependent, we here follow \citet{bathiany_detecting_2013-1} and set it to a constant value (here: $\tau = 100$ years), which allows a more straightforward interpretation of CSD in the model.
The constant $\tau$ resembles the inherent timescale of the vegetation system. 
The time step $\Delta t = 1$ corresponds to one year in accordance with $P$ representing mean annual precipitation.
While this contradicts the VOD observations that are available for  ~10 years, with fluctuations happening on monthly scale, it must be pointed out here that the time scale difference does not matter as a time step unit could be interpreted as e.g. days as well.

Realizations of the model are the numerical approximations of the solutions of the given equations by an Euler-Maruyama-scheme with random white noise added to the precipitation.
As this analysis focuses on measures that are valid in the vicinity of the fixed point, equilibrium runs are performed. For each change in the bifurcation parameter, 1000 steps of the Euler-Maruyama-scheme are executed, of  which the last one is added to the data set as one time step. Moreover, the data of the first time step are the result of 10000 steps with initial values randomly distributed around the fixed point.
The simulations shown are based on 1000 time steps.
All results are based on 1000 realizations of the stochastic model.

\subsection*{Resilience analysis}
\label{sec:methods_resilience_analysis}

To assess changes in resilience based on the concept of CSD, we de-trend and de-season the data sets and calculate the variance, AC1 and spatial correlation in sliding windows. The change is then defined as the time series' linear trends.

\subsubsection*{De-trending and de-seasoning}
\label{sec:methods_detrending}
All indicators of CSD are based on perturbations of the state variable around its equilibrium. Hence, the variable to analyze must not contain any trend or seasonality. 
For the conceptual model, this is achieved by subtracting the equilibrium vegetation state of the corresponding cell from its vegetation state in each realization.
For the satellite vegetation data, the trend and seasonality of each VI in each grid cell are removed by applying the Seasonal-Trend decomposition using LOESS (STL). The parameters are set to the default values proposed in \citet{cleveland_stl_1990}, but an additional analysis confirms that the general results are robust against variation in the parametrization.

\subsubsection*{Detection of resilience changes}
\label{sec:methods_CSD_indicators}
The actual indicators of resilience are then computed on sliding windows. For each window, variance and AC1 are calculated grid-cell-wise, with correlation referring to the Pearson correlation coefficient throughout this study.
The spatial correlation of a cell $i$ is defined as
\begin{linenomath*}
	\begin{equation*}
		C_i = \frac1n \sum_{j}{ Cor(r_i, r_j) } \qquad \text{for} \qquad j \in \Omega_i,
	\end{equation*}
\end{linenomath*}
where $r_i$ and $r_j$ are the residuals of the VI in the grid cells $i$ and $j$ on the current time window and $\Omega_i = \{j \| dist(i,j) \leq 100\text{km}\}$ and $n$ being the size of $\Omega_i$.
To put the formula into words, the spatial correlation is the mean of the cell's temporal correlation with all its neighbors, where neighbors are limited to all cells within a given radius. 
For the conceptual model, the radius is set to 1 and the distance between two cells $i$ and $j$ is defined as $\|i-j\|$. Hence, only directly neighboring cells are considered when computing a cell's spatial correlation value.
Regarding the VIs, the radius is 100\,km for the main results, where all distances are calculated based on the great circle distance between grid cell centers. However the results are robust against other choices of the radius.

For the conceptual model, the size of the windows is set to 100 time steps.
In the real data, defining the sliding window size must be done with consideration of the total length of the time series. Large window sizes ensure robust indicator values, but leave less time steps to calculate their corresponding trends or tendencies.
In the main analysis, the window size is 5 years, equivalent to 60 data points. The trade-off is especially high in the case of the single-sensor data considered here due to their short availability.
Yet, the results are robust with regards to the window size.

To evaluate the change in resilience over time, the development of the corresponding indicators of CSD over time is of interest. 
This change over time is then quantified by the trend, which is defined as the slope of a linear regression. 
While the trend might be biased by large jumps in comparison to many small changes, the Kendall $\tau$ correlation coefficient assesses the time series' tendency, measured as the steadiness of the increase. All results are stated in terms of trends, but are robust when compared to results assessed by the tendency.

\subsubsection*{Time of emergence}
\label{sec:methods:VECODE_time_of_emergence}
To assess the reliability of the indicators, we define the time of emergence (ToE) as the time a significant trend emerges for the first time.
To calculate the ToE for the conceptual model, the significance in terms of the p-value of each single trend in each time step is calculated. To do so, 1000 phase surrogates of each single realization's vegetation residual are created. For each surrogate, the indicators are then calculated in the same manner as for the actual state variable. 
For each time step, the trend in the surrogate's as well as original state variable's indicators until this point in time is measured. The p-value is then defined as the one minus the percentile in which the later is when assuming it results from the distribution given by the former.
An indicator at a given time and cell is regarded significant, if $p < 0.05$.
The time of emergence is then defined as the first time when the trend becomes significant.
The ToE was similarly defined e.g. in \citet{hawkins_time_2012} as the first time a signal-to-noise ratio is above a certain threshold.
An indicator of CSD used as an Early Warning Sign (EWS) should preferably warn as early as possible. Thus, indicators are better when their increase is significant long before a Tipping Point.
Furthermore, they are more reliable when the spread in the ToE over the realizations is small.

\section*{Data}
\label{sec:methods_satellite_data_preparation}

\begin{figure}
	\centering
	\includegraphics[width=.6\linewidth]{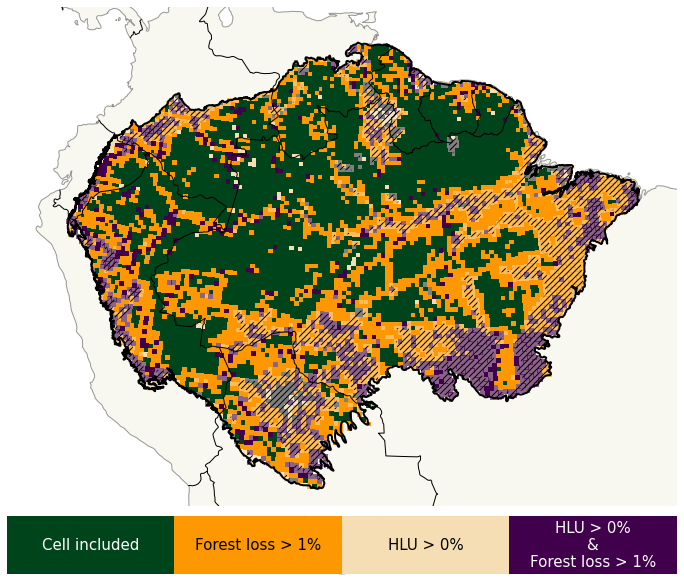}
	
	\caption{
		\textbf{Grid cells included in the analysis.}
		For a grid cell to be included (green), there may not be any remotely sensed Human Land Use \citep{friedl_mcd12c1_2015} (HLU, violet) and the forest loss according to \citet{hansen_high-resolution_2013} (orange) may not cumulate to more than $1\,\%$ of the cell's area over the years 2001 to 2020.
		Furthermore, the Evergreen Broadleaf Fraction (EBlF, hatched) may not be less than $80\,\%$ in any of the years, assuring that only dense rainforest is considered.
	}
	\label{fig:cells_used}
\end{figure}

Remote sensing of high biomass regions such as the ARF is challenging for several reasons. Vegetation Indices (VIs) based on optical imaging may fail due to the dense canopy that can lead to asymptotic saturation \citep{huete_overview_2002}.
Moreover, artifacts from persistent cloud cover and aerosols may remain in the processed VI \citep{samanta_why_2012}.
In contrast, VOD ~\citep{jackson_vegetation_1991,vreugdenhil_analyzing_2016} is derived from microwave satellite observations and linked to vegetation water content \citep{kirdiashev_microwave_1979} via which it can be interpreted as an indicator of canopy density and above-ground-biomass.
\citet{smith_empirical_2022} showed that VOD is more suitable for a vegetation resilience analysis based on CSD at the global scale.
Besides the reliability of the VI, sufficiently long time series are crucial for the analysis of the evolution of vegetation resilience. 
Yet, while long-time scale merged VOD products exist (e.g. VODCA \citep{moesinger_global_2020}), intercalibration techniques in multi-sensor observational products providing long time series can cause artifacts in any resilience analysis that is based on CSD \citep{smith_reliability_2023}. Hence, here we analyse single-sensor data, all of which was recorded over a time span of at least eight years.  
For the time period from 2000 to 2020 two sensors recorded suitable data.
First, the Advanced Microwave Scanning Radiometer Earth Observing System sensor (AMSR-E) \citep{vrije_universiteit_amsterdam_richard_de_jeu_and_nasa_gsfc_manfred_owe_amsr-eaqua_2011}  was active from June 2002 to October 2011. We combine the daily into monthly data by taking averages over full months following \citep{boulton_pronounced_2022}, so only the time period from July 2002 to September 2011 is analyzed. 
AMSR-E provides VOD data based on the C- as well as the X-band. The C- and X-bands stem from a sampling of frequencies around 6.9 and 10.7\,GHz., respectively.
In theory, shorter wavelengths (X-band) are mainly responsive to the moisture content of the canopy \citep{chaparro_l-band_2018,tian_coupling_2018,fan_evaluation_2018,konings_macro_2019}, while longer wavelengths (C-band) are more sensitive to deeper vegetation layers, including the woody parts \citep{andela_global_2013}.
AMSR-E's successor AMSR2 \citep{vrije_universiteit_amsterdam_richard_de_jeu_and_nasa_gsfc_manfred_owe_amsr2gcom-w1_2014} 
was launched in July 2012 and is still active. In this sensor the C-band was divided into two frequency bands termed C1 and C2 (6.9 and 7.3\,GHz), and VOD data can be derived from both.
The first complete month of AMSR2 is August 2012, and due to the availability of the data used to exclude Human Land Use (HLU), AMSR2 is analyzed until December 2020.

To access the changes in the Amazon rainforest  over the last decades, we use the VOD data derived from the nighttime overpass recordings as they are more suitable for VOD \citep{de_jeu_global_2008}, and the spatial resolution is kept at 0.25$^\circ$.


\subsection*{Grid cell selection}
\label{sec:methods_cells_used}

To define natural rainforest grid cells within the Amazon basin (\url{http://worldmap. harvard.edu/data/geonode:amapoly_ivb}), two datasets are used to check for the following two requirements. 
A cell must have at least $80\,\%$ evergreen broadleaf fraction (EBlF) and $0\,\%$ human land use (HLU), extracted from the MODIS Land Cover dataset \citep{friedl_mcd12c1_2015} (MCD12C1, Version 6), based on Land Cover Type 1 in percent. This data is available from 2001 until 2020. We define HLU as any of \textit{Croplands}, \textit{Urban and Built-up Lands}, and \textit{Cropland/Natural Vegetation Mosaics}.
To be as rigorous and conservative as possible, also any cell with more than one percent forest loss during the same period, 2001-2020, according to \citet{hansen_high-resolution_2013} is excluded. The percentage refers to the accumulated area of forest loss detected over time.
This data selection is visualized in Fig. \ref{fig:cells_used}.


\section*{Results}\label{sec:results}

\subsection*{Resilience indicators in a conceptual model of vegetation-atmosphere moisture recycling}
\label{sec:results:VECODE}

\begin{figure}
	\centering
	\includegraphics[width=.8\textwidth]{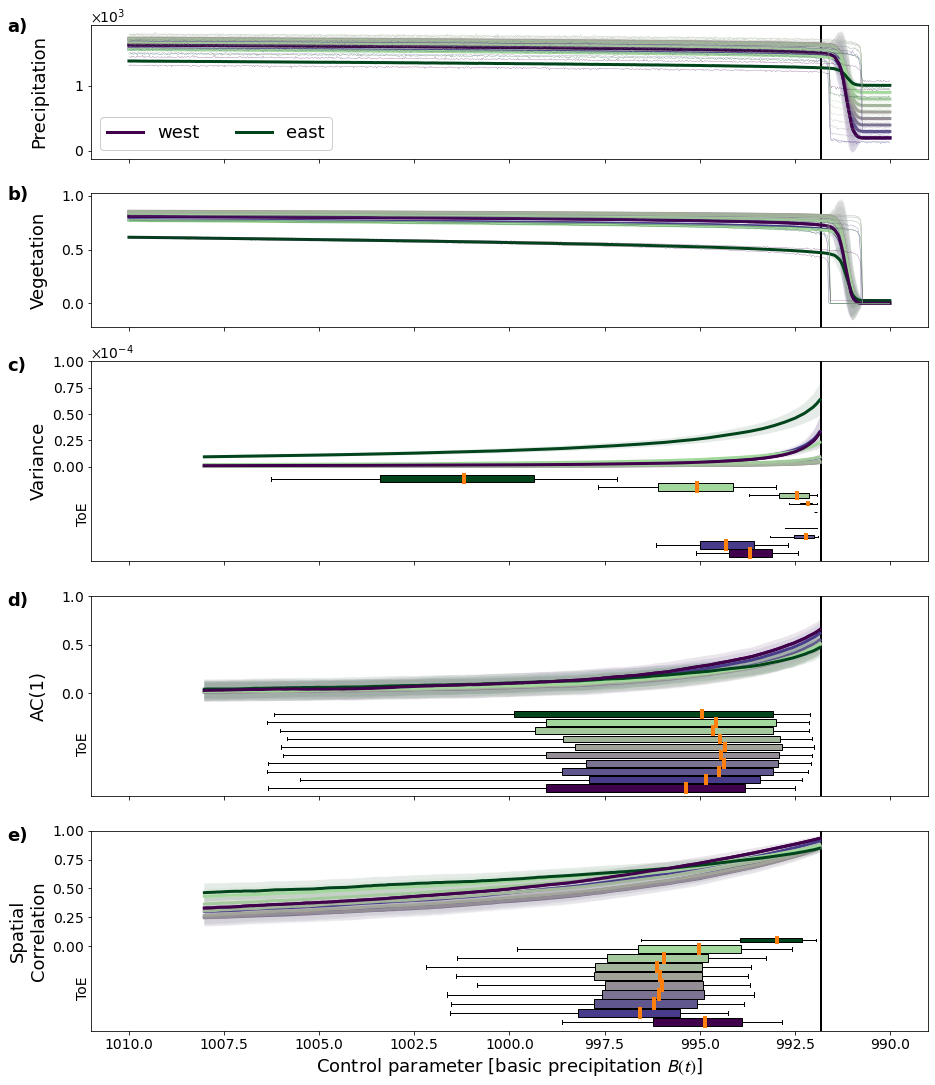}
	\caption{
		\textbf{Summary of 1000 realizations of the conceptual model.}
		Time series of \textbf{a)} precipitation, \textbf{b)} vegetation, \textbf{c)} variance, \textbf{d)} autocorrelation at lag 1 (AC(1)), and \textbf{e)} spatial correlation.
		Colors encode the 10 cells as in Fig. \ref{fig:VECODE_scheme} with violet being the western-most and green the eastern-most cell.
		Solid lines show the mean of the corresponding quantity.   
		For precipitation and vegetation, thin solid lines mark the maximum and minimum over all runs at each time step.
		For the indicators of CSD, shaded bands mark one standard deviation around the mean.
		The vertical black lines mark when the total precipitation of at least one cell in at least one realization drops below the critical value for the first time. 
		The boxplots in \textbf{c-e)} display the distribution of the time of emergence (ToE, when the corresponding indicator's trend becomes significant, see Methods section). 
		They extend from the first to the third quartile, with the orange line marking the median. The whiskers extend from the 5th to the 95th percentile.
		If for a given cell a realization's trend did not become significant at least once, the height of the box is decreased. This is the case for the variance of several cells as well as e.g. the spatial correlation of the eastern-most cell.
		Note that due to the linear decrease of $B(t)$, the control parameter on the x-axis can be directly linked to time.
	}
	\label{fig:VECODE_results}
	
\end{figure}

The conceptual model implements moisture-recycling from east to west in combination with decreasing moisture inflow from the Atlantic ocean. The latter one corresponds to a potential climate change scenario and acts as the forcing in the model. 
Once the point where the eastern cell tips to a low vegetation state is reached, the abrupt decline in vegetation  reduces the inflow of recycled moisture in all the cells further to the west, thus inducing a tipping cascade.
It is important to note that within one realization, all cells tip simultaneously, so the spread in the actual tipping point in Fig. \ref{fig:VECODE_results}b is a result of stochastic differences between the realizations. 
Before the Tipping Point, the resulting decline in precipitation is almost linear for all of the cells, compare the zoom-in into precipitation in Fig. \ref{fig:SI_VECODE_precipitation}. 
Interestingly, deforestation in the east-most cell mimicked by a artificial linear reduction of this cell's vegetation state also induces a cascade of precipitation and vegetation tipping (see Fig. \ref{fig:SI_VECODE_results_defo}).

In Fig. \ref{fig:VECODE_results}, the vertical black lines mark the time when the total precipitation of at least one cell in at least one realization drops below the critical value for the first time. 
Thus, only time steps before entering this region are used in the resilience analysis.
For all cells, variance, AC1 and spatial correlation as indicators of CSD are calculated on sliding windows and their change is assessed as the linear trends. For all indicators and all cells, the three indicators increase on average with almost no negative trends (see Fig. \ref{fig:SI_VECODE_results_violins}).

As defined in the Methods Section, the time of emergence (ToE) is the time when the trend of an indicator becomes significant for the first time.
In Fig. \ref{fig:VECODE_results} the median ToE is highlighted by an orange line and differs both between different cells and between different indicators for the same cell. 
For the eastern cell, variance is the first indicator to become significant. This early emergence is because this cell is the closest to a standalone tipping point, and so its variance is increasing towards infinity and thus the increase becomes significant earlier in time. This earlier increase shows that the tipping of the eastern cell triggers the cascading tipping of the other cells.
Note that for this cell, the spatial correlation does not become significant in all realizations.
For the second easternmost cell, the median ToE is comparable between the spatial correlation and the variance.
For all other cells except the westernmost the spatial correlation outperforms the other indicators as an EWS (for the westernmost cell the AC1 has the earliest ToE). The comparison to the variance is especially striking, as the variance trend's significance emerges very late or not at all for these cells.
One can thus conclude that the AC1 and spatial correlation become significant earlier and more often for the coupled cells, with spatial correlation being ahead in time for most cells.  On the other hand, variance exhibits early significant trends for the (almost) uncoupled cells.
Furthermore, an indicator is regarded more reliable when its trend consistently becomes significant at one time. This can be translated to a smaller spread in its ToE. 
It is apparent that the spread in ToE is smaller for spatial correlation for the highly coupled cells, indicating its robustness in systems with strong coupling. 
However, any indicator can sometimes have spuriously significant trends that lose their significance at a later time, so it is informative to also define a `permanent' ToE. This ToE is defined as the time when the emergence of a significant trend is permanent for the rest of the time period, and is shown in Fig. \ref{fig:SI_VECODE_results_psigpermanent}. 
While this definition reduces the large uncertainty in the ToE of AC1, the choice of the method does not alter relations between the median ToEs of the different indicators. The overall result also stays the same: the spatial correlation is the most reliable and earliest indicator to have significant positive trends in coupled cells.

We conclude that, whilst their significance might emerge at different times, all three indicators are expected to increase for spatially extended systems like the ARF when approaching a critical transition. 
Moreover, the comparatively early and tightly spread time of emergence of the spatial correlation proves its reliability as an indicator of CSD in coupled systems.
Consequently, if the ARF is losing resilience we would expect coherent and significant increases in all three indicators in observations of Amazon vegetation, with most pronounced signs in the spatial correlation.

\subsection*{Vegetation resilience in the Amazon basin}

\begin{figure}
	\centering
	\begin{minipage}{\textwidth}
		\includegraphics[width=\linewidth]{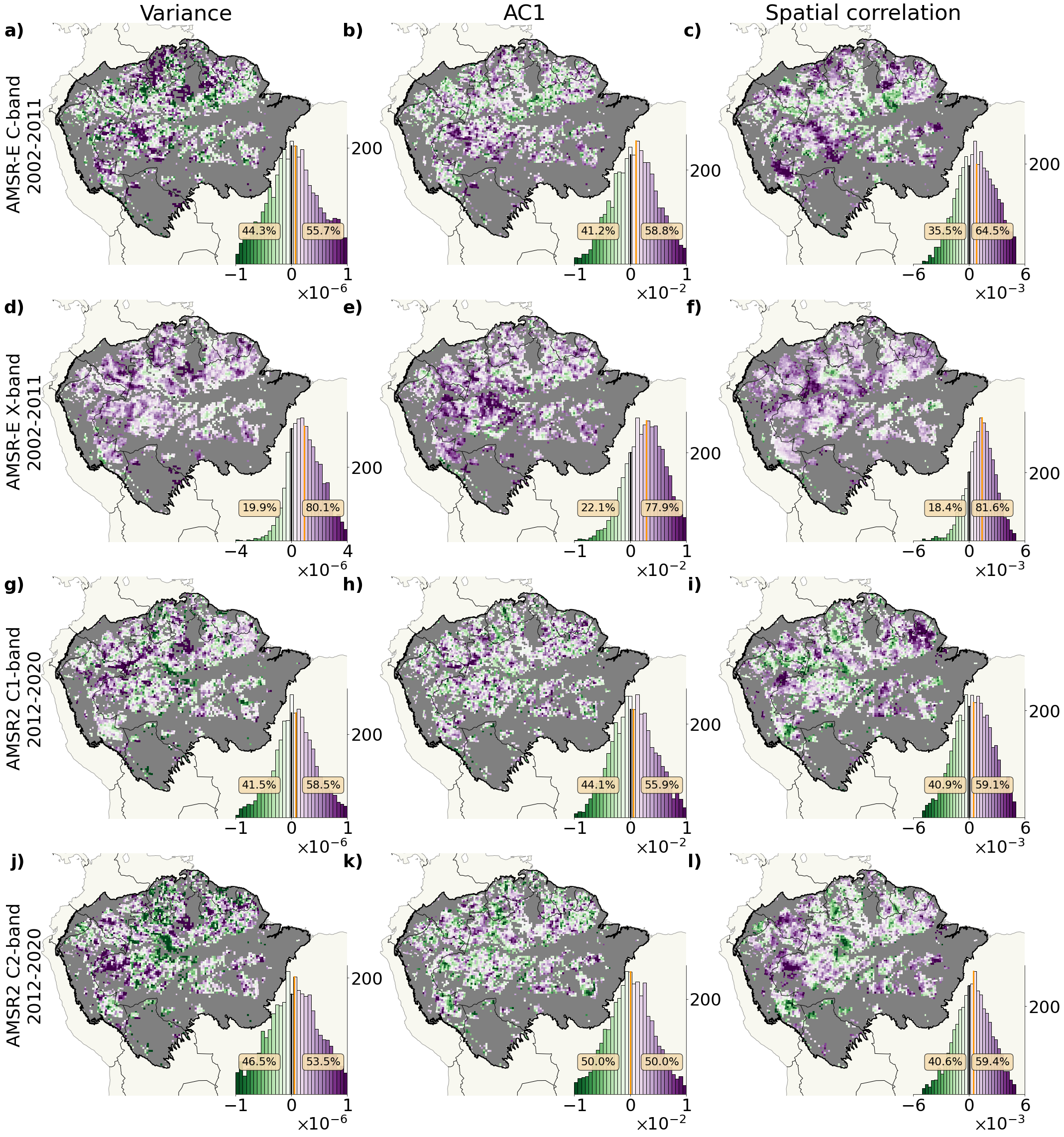}
	\end{minipage}%
	
	\caption{
		\textbf{Changes in the different CSD indicators for the different Vegetation Optical Depth data sets.}
		The change is assessed by linear trends of \textbf{a,\,d,\,g,\,j)} variance, \textbf{b,\,e,\,h,\,k)} AC1 and \textbf{c,\,f,\,i,\,l)} spatial correlation as distributed across the whole Amazon basin based on \textbf{a-c)} AMSR-E's C-band and \textbf{d-f)} X-band for the years 2002 to 2011 and \textbf{g-i)} AMSR2's C1-band and \textbf{j-l)} and C2-band.
		The inlay histograms visualize the distribution of the trends with the same color-coding as on the maps. 
		There, zero is marked by a black vertical line and the median of the distribution by an orange vertical line.
		Note that the colors and bins are restricted to suitable values.
	}
	\label{fig:AMSRE_indicators}
\end{figure}

\begin{figure}
	\centering
	\begin{minipage}{\textwidth}
		\includegraphics[width=\linewidth]{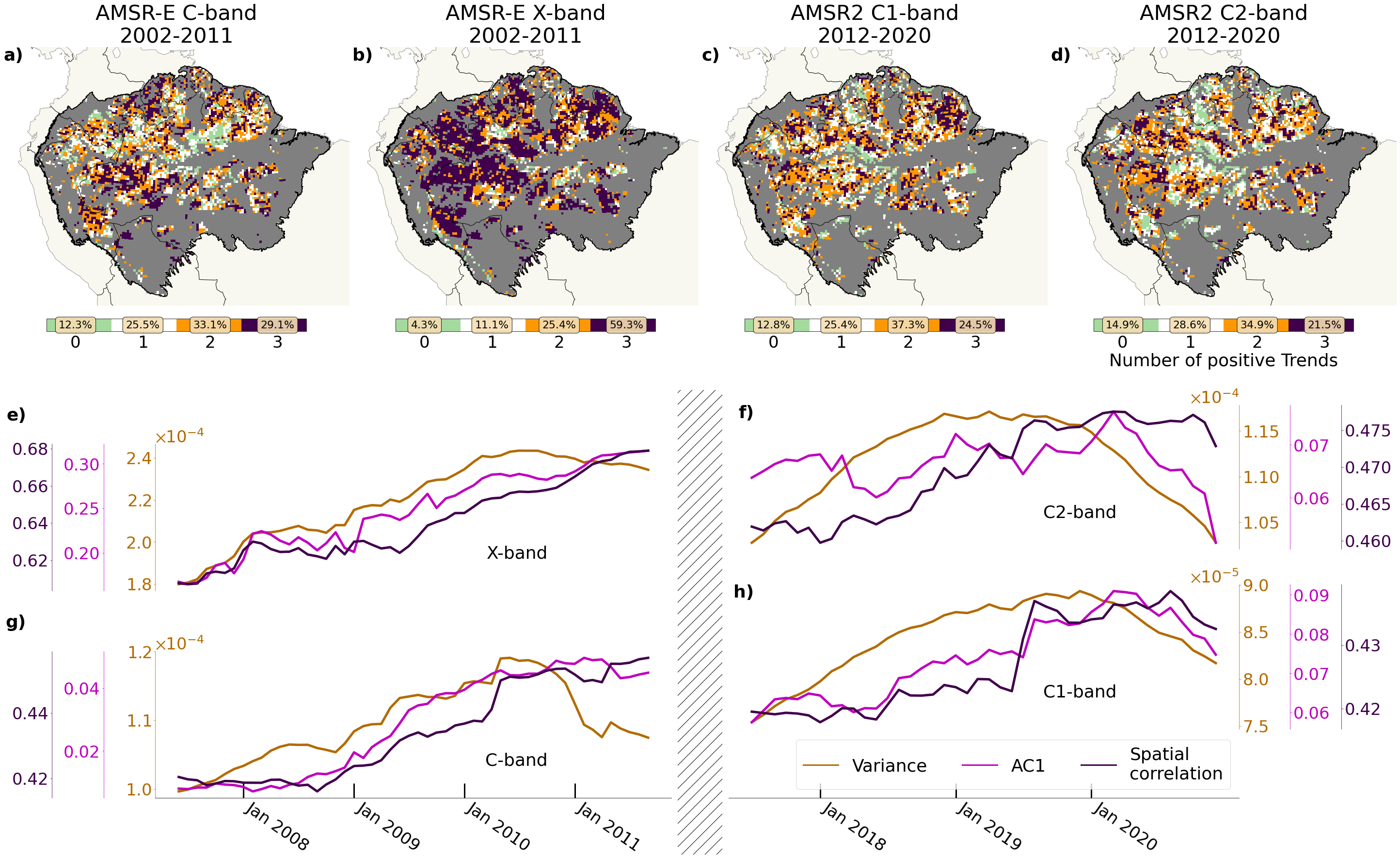}
	\end{minipage}%
	
	\caption{
		\textbf{Resilience changes detected in the different Vegetation Optical Depth data sets.}
		The assessment is based on \textbf{a,\,g)} AMSR-E's C-band and \textbf{b\,e)} X-band as well as \textbf{c,\,h)} AMSR2's C1-band and \textbf{d,\,f)} C2-band.
		The maps in \textbf{a-d)} summarize the results for the different CSD indicators by displaying the number of increasing indicators per grid cell, with the percentage of co-occurrences denoted below.
		The time series in \textbf{e-h)} show the spatially averaged changes for each CSD indicator over the respective time period, allowing for an estimation of the overall resilience change of the vegetation in the Amazon basin as one ecosystem.
	}
	\label{fig:AMSRE_summary}
\end{figure}

As the ARF is a spatially coupled system, similar to the simple moisture recycling model presented above, we expect to see increasing indicators of CSD in case of resilience loss.
Three indicators are considered here, namely variance, AC1, and spatial correlation. The change in an indicator's time series is quantified by the linear trend (see Methods section). 
Fig. \ref{fig:AMSRE_indicators} displays the spatial pattern of trends in the individual indicators as found for the different data sets.
It is important to keep the different time spans of the two sensors in mind, as the results can only be compared within each sensor's bands.

\textbf{2002 - 2011 (AMSR-E):}
From July 2002 to September 2011, AMSR-E was active. The changes in the single indicators for AMSR-E's C- and X-band are depicted in Fig. \ref{fig:AMSRE_indicators}a)-f).
For its C-band, the distributions of trends of all indicators have a positive median, so overall more cells exhibit a positive trend than not. 
Whilst the spatial distributions of trends in variance, AC1 and spatial correlation are distinct, in all three the positive trends cluster in the southwest and along the northern basin boundary.
AMSR-E's X-band has, for all three indicators, an even stronger tendency towards cells with a positive trend.
The regions of positive trends comprise those of the C-band, but extend across the whole Amazon basin, with the only exceptions being some cells with non-positive trends in the east, close to the Amazon river.

As explained above, in the case of CSD we would expect the changes in all three indicators to be coherent.
Fig. \ref{fig:AMSRE_summary} summarizes the results from the three indicators. The maps show the number of indicators exhibiting a positive trend at each grid cell. 
For {AMSR-E's} C-band, Fig. \ref{fig:AMSRE_summary}a) confirms that the most prominent patch of positive trends in all indicators is the southwestern part of the Amazon basin. Further signs of extended resilience loss are noticeable along the northern basin boundary.
For the X-band, the overall resilience loss of the ARF, yet especially pronounced in the southwest, becomes apparent in Fig. \ref{fig:AMSRE_summary}b and e). 
The time series of spatial averages suggest that the ARF as an interacting ecosystem has, on average, lost resilience over the years 2002 to 2011, with clear signals in both bands.

\textbf{2012 - 2020 (AMSR2):}
Turning to the time period from August 2012 until December 2020, for which VOD data based on AMSR2 was analyzed, the signal is less clear (Fig. \ref{fig:AMSRE_indicators}).
Even so, except for the AC1 of the C2-band, the trends of all indicators are more often positive than negative.
As for AMSR-E, the two bands in AMSR2 have different spatial trend distributions.
For the C1-band, patches of positive trends, mostly in the variance and spatial correlation, are visible in the northeast as well as in the west.
For the C2-band, the signals are stronger in the west but less pronounced in the northeast, and again strongest in the variance and spatial correlation.
The spatial comparison of the trends in the three indicators in Fig. \ref{fig:AMSRE_summary}c) reveals that, based on the C1-band, the vegetation in the northeast of the Amazon basin 
has lost resilience. Even though no clear signals in this region were found for the years 2002-2011 based on AMSR-E's C-band, its X-band does show destabilization in this region already in the years before 2012.
Considering the C2-band, positive trends co-occur mainly in the very west, where both bands from AMSR-E indicate a loss of resilience for the years before.
The spatially averaged indicator time series clearly increase over the period until around mid-2019. Interestingly, from then on until the end of the study period, all time series decrease, although the decrease in the spatial correlation is marginal compared to the previous increases. 
We restrict our study period to years before 2020 as the data for excluding human land use is not available thereafter, but using a less conservative analysis we find that the indicators continue to decrease in the 2020-22 period. Yet, this decrease could be explained by forest loss and the results for AMSR2 analyzed until 2022 cannot be relied on (see Fig. \ref{fig:SI_AMSR2_2022}).
Furthermore, AMSR2 also provides an X-band (10.7\,GHz), for which results are shown in the supplement (Figs.~\ref{fig:SI_AMSR2_bandX_1} and \ref{fig:SI_AMSR2_bandX_2}) as the theory implies that it is less suitable for biomass assessment in the Amazon than the C-bands.

All results are robust with respect to the parametrization of the de-trending method STL (see Figs.~\ref{fig:SI_STL} and \ref{fig:SI_STL_summary}), the size of the sliding windows (see Figs.~\ref{fig:SI_sws} and \ref{fig:SI_sws_summary}), the measure of increase in the indicator time series (see Figs.~\ref{fig:SI_KendallTau} and \ref{fig:SI_KendallTau_summary}), and the maximum distance that defines `neighbors' (see Fig. \ref{fig:SI_distances}).

Even though they reveal a less dramatic picture of the condition of the Amazonian rainforest than that found by \citet{boulton_pronounced_2022}, these results confirm the resilience loss found in that work, which was based on the AC1 indicator and the merged data set VODCA and started in the early 2000s.
In this work, we use a number of different indicators of CSD as well as different observations, which makes our results especially robust.
In particular, for all data sets considered the number of cells where all three indicators show a positive trend is more than double the expected number of 12.5\,\% ($=0.5^3$, with 29.5\,\% and 59.3\,\% for AMSR-E's C- and X-band and 24.5\,\% and 21.5\,\% for AMSR2's C1- and C2-band, respectively, see also Fig. \ref{fig:AMSRE_summary}a-d)).

\section*{Conclusions}\label{sec:discussion}
In spatially coupled systems, the spatial correlation is expected to increase prior to a critical transition, establishing an indicator of CSD.
In this work we first used a conceptual model to show that for a system with spatial extension and coupling similar to the ARF, variance, AC1 and spatial correlation increase as it approaches a critical transition. This is the case even if a cascade of tipping is induced by a single cell \citep{bathiany_detecting_2013-1}. The simulations revealed that for strongly coupled cells where a transition is caused by a reduction of the incoming recycled moisture, spatial correlation is an especially reliable and early mean of detecting the loss of resilience and an approaching transition.

Recent studies have shown that satellite data are appropriate only under certain conditions for investigating changes in the resilience of the ARF \citep{smith_empirical_2022,smith_reliability_2023}. In particular, it has been shown that time series which combine different data sources might inherit artifacts resulting from the merging procedure \citep{smith_reliability_2023}, and thus in our work we exclusively analyzed single sensor data.

While time series of several decades would be favorable to capture long-term vegetation dynamics, shorter time series are still capable of sensing physiological responses to droughts and other environmental conditions.
The sensors AMSR-E and AMSR2 provide acceptably long VOD time series (2002-2011 and 2012-2020, respectively), which we analyzed on a monthly resolution, following \citet{boulton_pronounced_2022}.
For the early 2000s we find an overall increase of the CSD indicators, with more striking signs of resilience loss in AMSR-E's X-band. The spatial pattern is consistent across the two bands, with the largest losses of resilience occurring in the southwest and north.
From 2012 to 2020, AMSR2 data reveals a less clear picture. Yet, the cells in the C1-band where all three indicators increase reside mostly in the northeast, coinciding with the resilience loss detected by AMSR-E in the preceding years. The cells that are likely undergoing destabilization according to AMSR2's C2-band are concentrated in the southwest.

Overall, even though the results differ somewhat for the individual data sets, we can conclude that the ARF's vegetation experienced a loss of resilience during the first two decades of the 21st century. More pronounced signals were found for the time period from 2002 to 2011, with the regions of destabilization comprising the western Amazon basin, the band along the northern boundary as well as the northeastern parts.
Interestingly, the regions in the southwest where destabilization is detected in all data sets correspond to the regions downstream from the `atmospheric rivers'. Hence, they are highly dependent on moisture recycling, implying that they are considerably spatially coupled and found to be more vulnerable to tipping due to network effects \citep{wunderling_recurrent_2022}.
In combination with the results from the conceptual model, which show that spatial correlation gives an especially reliable indication of CSD in a highly coupled sub-system, the spatial correlation can be considered the most reliable indicator in the southwest. This is in line with the fact that all data sets show increasing spatial correlation in parts of the southwestern Amazon. These increases could hint at a destabilization due to changes in incoming recycled moisture, which in return could be an effect of the high deforestation rates in the `arc of deforestation' further upstream of the `atmospheric rivers'.

The main forcing affecting the ARF's vegetation and potential resilience changes is presumably the precipitation.
Thus, it is essential to ensure that the detected changes in the vegetation's indicators of CSD are not a direct representation of corresponding changes in precipitation statistics. We thus calculate the same CSD indicators for the precipitation at each grid cell, and analyse the regions in which the sign of their trends agree with the sign of the trend in VOD CSD indicators (see Fig. \ref{fig:SI_VegPrecip_CSD_indicators}). This analysis shows that the signals found in Fig. \ref{fig:AMSRE_summary} are not a consequence of statistical changes in precipitation.
On the other hand, it is still possible that the vegetation resilience loss is related to a decrease in the annual precipitation. Yet Fig. \ref{fig:SI_precipitation_attribution} shows that the signs of CSD cannot always be explained by negative trends in precipitation. However, this could be caused by a lag in time between the changes in precipitation and the resilience loss in vegetation. 
Furthermore, the lack of a negative trend  in mean annual precipitation could still coincide with overall drying: in case of an increasing evapo-transpirative demand or, as many studies have suggested, shifts in the dry season length and strength, and increasing frequency of droughts that could be evened out by increases in precipitation during the rainy season or flood years \citep{marengo_changes_2018}.
Thus, further work is needed to better understand the interplay of causes that can drive the ARF towards a dieback.

This work has focused on spatial correlation as an indicator of CSD, due to the spatially coupled structure of the ARF and the theoretical advantages this indicator shows in numerical experiments. 
Yet, multiple other potential (spatial) CSD indicators exist, such as spatial variance, spatial autocorrelation, spatial skewness \citep{dakos_slowing_2011,donangelo_early_2010,dakos_spatial_2010,guttal_spatial_2009}, or spatial permutation entropy \citep{tirabassi_entropy-based_2023} .
Several of these would be applicable in this setting, but their thorough investigation and comparison was beyond the scope of this study. Still, a comparison of different spatial resilience indicators could improve our understanding of their applicability as well as their reliability to detect changes in resilience in the ARF, as well as in other spatially coupled ecosystems.

If we are to robustly capture resilience changes, our efforts must be focused in a few key directions.
First, long single-sensor time series are preferred to reliably trace the dynamics and potential resilience changes of vegetation ecosystems.
Second, sensors and their derived VIs must be adequate to address the question of interest. To that end, it is important to find measures to assess the suitability of data sets. In the context of residuals-based resilience analyses, a sufficiently high signal-to-noise ratio is crucial. 
Furthermore, with respect to dense vegetation such as the ARF, it would be helpful to better understand problems induced by saturation in the VIs on their higher order statistics \citep{smith_reliability_2023}. 
Third, the applicability of indicators of CSD in different settings must be better understood, such that the most suitable approaches can be chosen depending on the system analyzed.

Overall, the complex changes we find in the ARF suggest that combining multiple datasets and indicators can give a clearer picture on the applicability of CSD, and the statistical robustness of trends in different parts of the Amazon.

Our findings suggest that the previously found loss of resilience in the early 2000s \citep{boulton_pronounced_2022,smith_empirical_2022} can, in parts, be confirmed by our approach and data, with less distinct signals for the years 2012 to 2020. Nevertheless, we find a destabilization of the vegetation in the ARF since the beginning of the century, independent of the data source or the indicator of resilience change, that is especially pronounced in the southwestern Amazon basin.

\section*{Open Research Section}
All data used in this study is publicly available.

For this study, only cells within the Amazon basin (\url{https://worldmap.maps.arcgis.com/home/item.html?id=f2c5f8762d1847fdbcc321716fb79e5a}, accessed on January 28, 2021) are considered. 
Human Land Use is extracted from the MODIS Land Cover dataset \citep{friedl_mcd12c1_2015} (MCD12C1, Version 6) available at \url{https://lpdaac.usgs.gov/products/mcd12c1v006/} (accessed on November 11, 2021), based on Land Cover Type 1 in percent. 
The Hansen deforestation data \citep{hansen_high-resolution_2013} was downloaded on May 31, 2022, from \url{https://storage.googleapis.com/earthenginepartners-hansen/GFC-2021-v1.9/download.html}.

The VOD from AMSR-E \citep{vrije_universiteit_amsterdam_richard_de_jeu_and_nasa_gsfc_manfred_owe_amsr-eaqua_2011}  (LPRM-AMSR\_E\_L3\_D\_SOILM3\_V002, C- and X-band) and AMSR2 \citep{vrije_universiteit_amsterdam_richard_de_jeu_and_nasa_gsfc_manfred_owe_amsr2gcom-w1_2014} (LPRM-AMSR2\_L3\_D\_SOILM3\_V001, C1- and C2-band) can be found at \url{https://hydro1.gesdisc.eosdis.nasa.gov/data/WAOB/LPRM_AMSRE_D_SOILM3.002} and \url{https://hydro1.gesdisc.eosdis.nasa.gov/data/WAOB/LPRM_AMSR2_D_SOILM3.001/} and were last accessed on November 8 and December 24, 2022, respectively.
The precipitation data from CHIRPS \citep{funk_climate_2015} is available at \url{https://data.chc.ucsb.edu/products/CHIRPS-2.0/global_monthly/netcdf/} and was last accessed on November 16, 2022. It was downscaled to the same resolution as the VOD data by selecting only the grid cells matching VOD's grid cell centers (center of $5\times 5$ cells).

\section*{Conflict of interest/Competing interests}
The authors declare no competing interests.

\section*{Author contributions}
LB, CB, and NB conceived and designed the research. LB conducted the analysis and prepared a first version of the manuscript. All authors discussed results, drew conclusions and edited the manuscript.

\section*{Supplementary information}
Supplementary figures accompany this paper in the supplementary material.

\section*{Acknowledgments}
	This is TiPES Contribution \#223; the Tipping Points in the Earth System (TiPES) project has received funding from the European Union's Horizon 2020 research and innovation programme under Grant Agreement No. 820970.
	Furthermore, this work has received funding from the VolkswagenStiftung, from the Marie Sklodowska-Curie grant agreement No. 956170, by the Federal Ministry of Education and Research under grant No. 01LS2001A, the DFG Grant SM 710/2-1, by the DARPA AI-assisted Climate Tipping point Modelling (ACTM) project (No. HR0011-22-9-0031) and the Bezos Earth Fund (No. XXX).

\bibliography{references}

\pagebreak
\appendix

\renewcommand{\thefigure}{S\arabic{figure}}
\setcounter{figure}{0}

\title{Supplementary Material\\for\\\ \\
	Spatial correlation  increase in single-sensor satellite data reveals loss of Amazon rainforest resilience}
\maketitle
\pagebreak

\section{Additional insight from the conceptual model}

\subsection*{Text S1. Stability in the conceptual model}
\begin{figure}
	\centering
	\includegraphics[width=\textwidth]{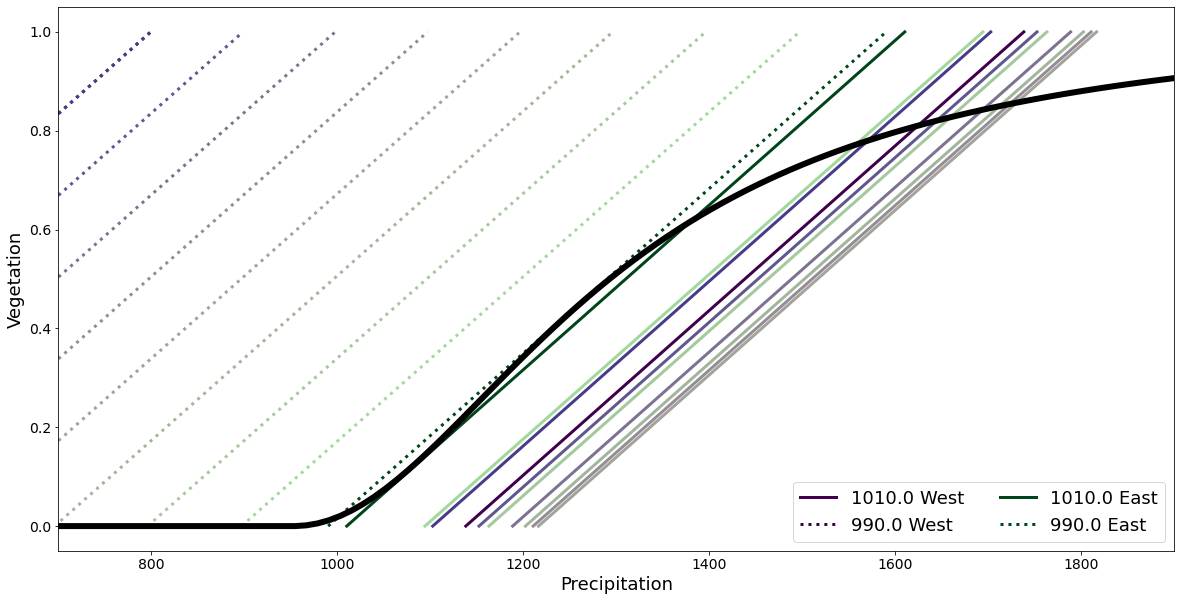}
	\caption{
		\textbf{Stability diagram of the conceptual model.}
		The interaction of vegetation and precipitation is plotted as two functions: vegetation as a function of precipitation and the other way around.
	}
	\label{fig:SI_VECODE_stability}
	
\end{figure}

The interaction of vegetation and precipitation is plotted in Figure \ref{fig:SI_VECODE_stability}.
In the conceputal model, vegetation is a function of absolute incoming precipitation only and is not dependent on the cell's location. Hence, it is the same for each cell and plotted as the single black solid line. 
Precipitation on the other hand is a function of the vegetation of the cell, but also of the vegetation of the cell's eastern neighbors. Hence, it differs for all cells, which are encoded by the colors as in Figure~\ref{fig:VECODE_scheme}. For the calculation of one cell's precipitation as a function of vegetation, its eastern neighbors are considered to be in equilibrium.
Furthermore, precipitation depends on the amount of moisture coming in from the Atlantic ocean, or in other words on the control parameter $B$.
The solid lines represent the state in the beginning of the experiment with $B = 1010$, and the dashed lines in the end with $B=990$.
Where the black and coloured lines cross at some $(V^*, P^*)$, the corresponding value of vegetation supports the value of precipitation $P^*$, but also is $V^*$ supported by $P^*$. Hence, this point is considered a fixed point, or equilibrium. 
As the moisture inflow is reduced in the simulations from 1010 (solid lines) to 990 (dashed lines), the fixed points with high vegetation values vanish, explaining the tipping in the  Figures~\ref{fig:SI_VECODE_results_violins} and~\ref{fig:SI_VECODE_results_psigpermanent} .

\subsection*{Text S2. Precipitation decrease in the conceptual model}
\begin{figure}
	\centering
	\includegraphics[width=\textwidth]{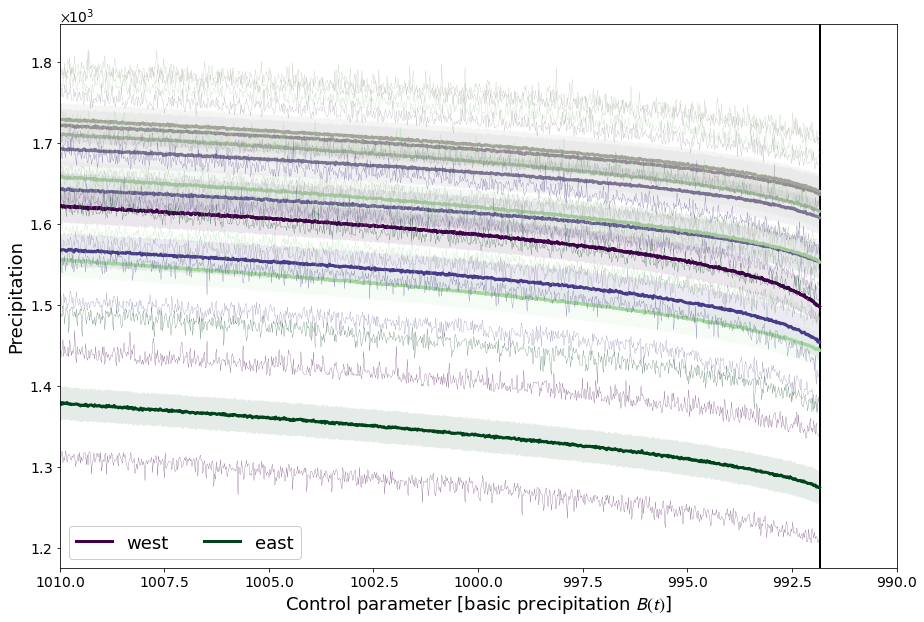}
	
	\caption{
		\textbf{Precipitation reduction in the conceptual model.}
		This Figure gives a zoom-in into cell-wise total precipitation, highlighting its stochasticity as well as the almost linear decrease up to the Tipping Point.}
	\label{fig:SI_VECODE_precipitation}
	
\end{figure}

The results of simulations with decreasing moisture inflow from the Atlantic ocean (decreasing $B(t)$) were presented in the main text in Figure~\ref{fig:VECODE_results}. Figure \ref{fig:SI_VECODE_precipitation} zooms in into cell-wise total precipitation and highlights its stochasticity as well as the almost linear decrease up to the Tipping Point of vegetation.

\subsection*{Text S3. Deforestation scenario leads to Tipping in the conceptual model}

\begin{figure}
	\centering
	\includegraphics[width=.8\textwidth]{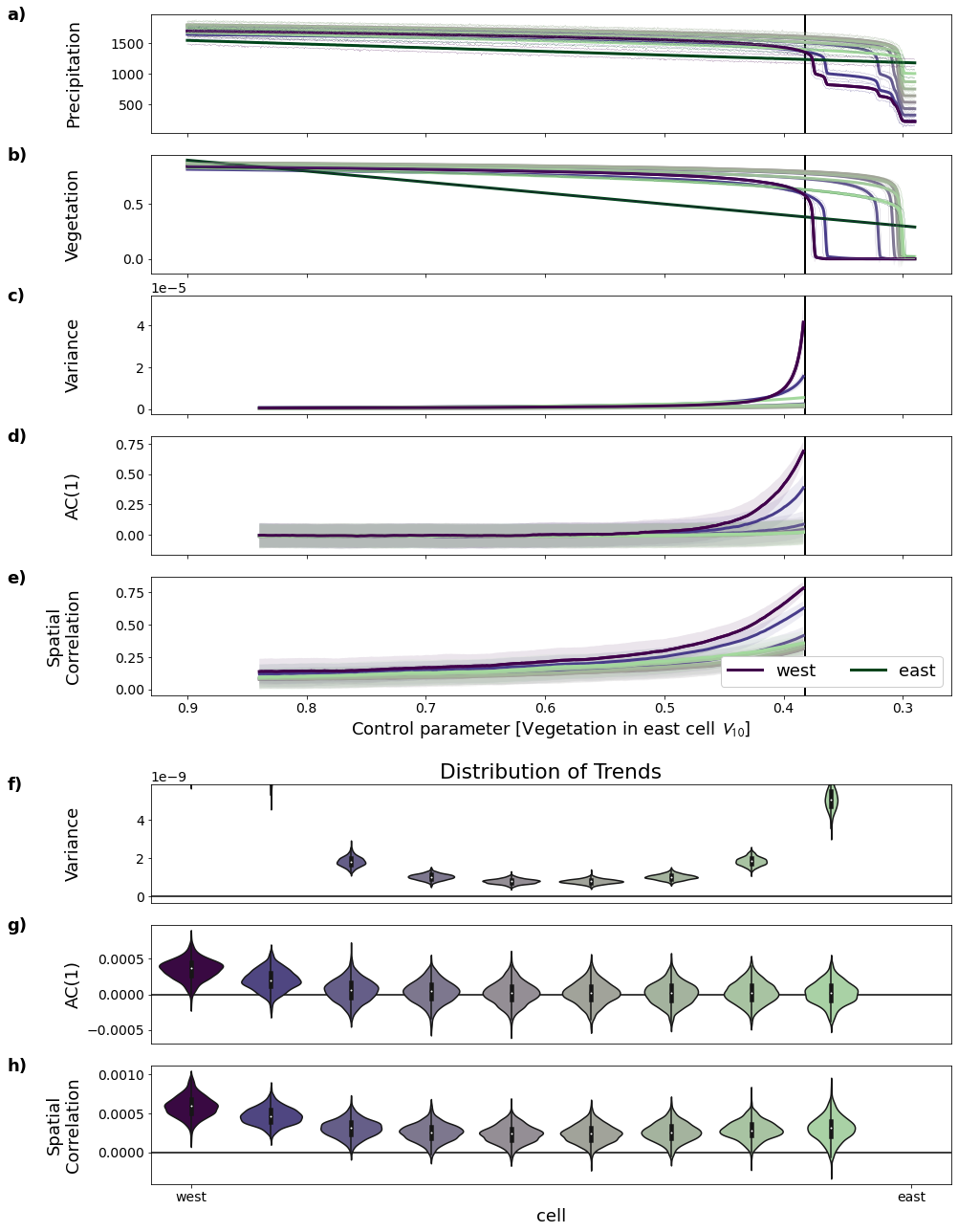}
	
	\caption{
		\textbf{Deforestation scenario of the conceptual model.}
		Tipping of vegetation can be achieved by constantly decreasing the vegetation in the east-most cell  from 1 to 0, thereby mimicking deforestation.
	}
	\label{fig:SI_VECODE_results_defo}
	
\end{figure}

The results of simulations with decreasing moisture inflow from the Atlantic ocean (decreasing $B(t)$) were presented in the main text in Figure~\ref{fig:VECODE_results}. Figure \ref{fig:SI_VECODE_results_defo} is identical but presents the results from simulations with unchanging inflow from the Atlantic, i.e. $B(t)=1000$ for all $t$. Instead, the amount vegetation in the east-most cell is considered the bifurcation parameter and constantly decreased from 1 to 0, thereby mimicking deforestation.

\subsection*{Text S4. Distribution of the trends for the three resilience indicators}
\begin{figure}
	\centering
	\includegraphics[width=\textwidth]{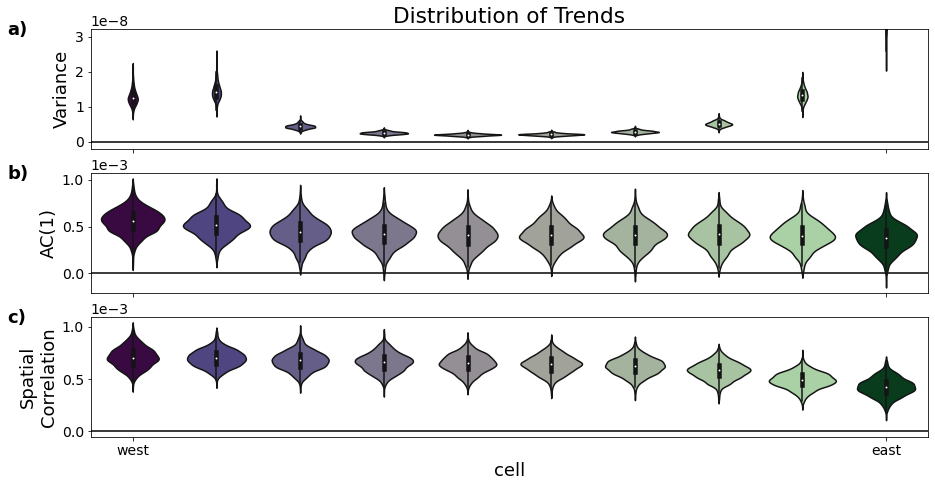}
	
	\caption{
		\textbf{Trends of the indicators of CSD in the conceptual model.}
		The violins represent the distribution of trends over 1000 realizations per cell for the three indicators variance, AC1, and spatial correlation.
	}
	\label{fig:SI_VECODE_results_violins}
	
\end{figure}

In the conceptual model, all indicators show clear trends when approaching the Tipping Point of vegetation, as illustrated in Figure \ref{fig:SI_VECODE_results_violins}.
There, clearly all three indicators show a strong shift towards positive values with only very few negative ones for AC1. In a spatially extended system like the ARF, one would thus expect all three indicators to increase when the corresponding grid cell loses resilience.

\subsection*{Text S5. Distribution of the trends for the three resilience indicators}
\begin{figure}
	\centering
	\includegraphics[width=.9\textwidth]{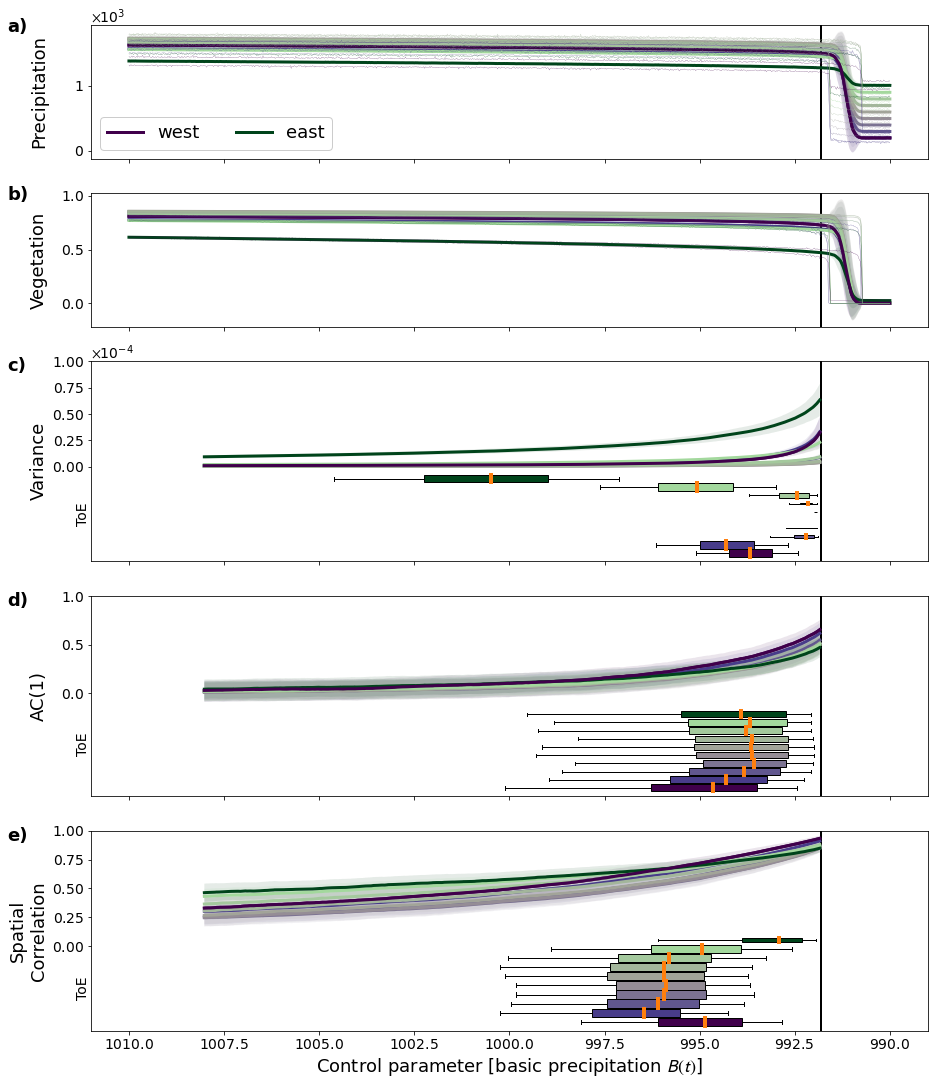}
	\caption{
		\textbf{Time of emergence of the indicators.}
		This is equivalent to Figure~\ref{fig:VECODE_results} but with the time of emergence defined as the time when a trend becomes \textit{permanently} significant.
	}
	\label{fig:SI_VECODE_results_psigpermanent}
	
\end{figure}

In Figure~\ref{fig:VECODE_results} the time of emergence (ToE) was defined as the first time when a trend becomes significant. Yet, as more conservative approach one can define the ToE as the time when a trend becomes \textit{permanently} significant, i.e. when all trends later in time are significant as well. The results with this definition are displayed in Figure \ref{fig:SI_VECODE_results_psigpermanent}.
The uncertainty in the timing of this type of ToE differs much less for AC1 compared to the definition in the main text.
This indicates that AC1 might be more prone to false alarms.

\FloatBarrier
\clearpage

\section{AMSR2 until 2022 excluded due to missing Human Land Use data}

\subsection*{Text S6. The influence of forest loss for the years 2021 to 2022}
\begin{figure}
	\centering
	\includegraphics[width=.9\textwidth]{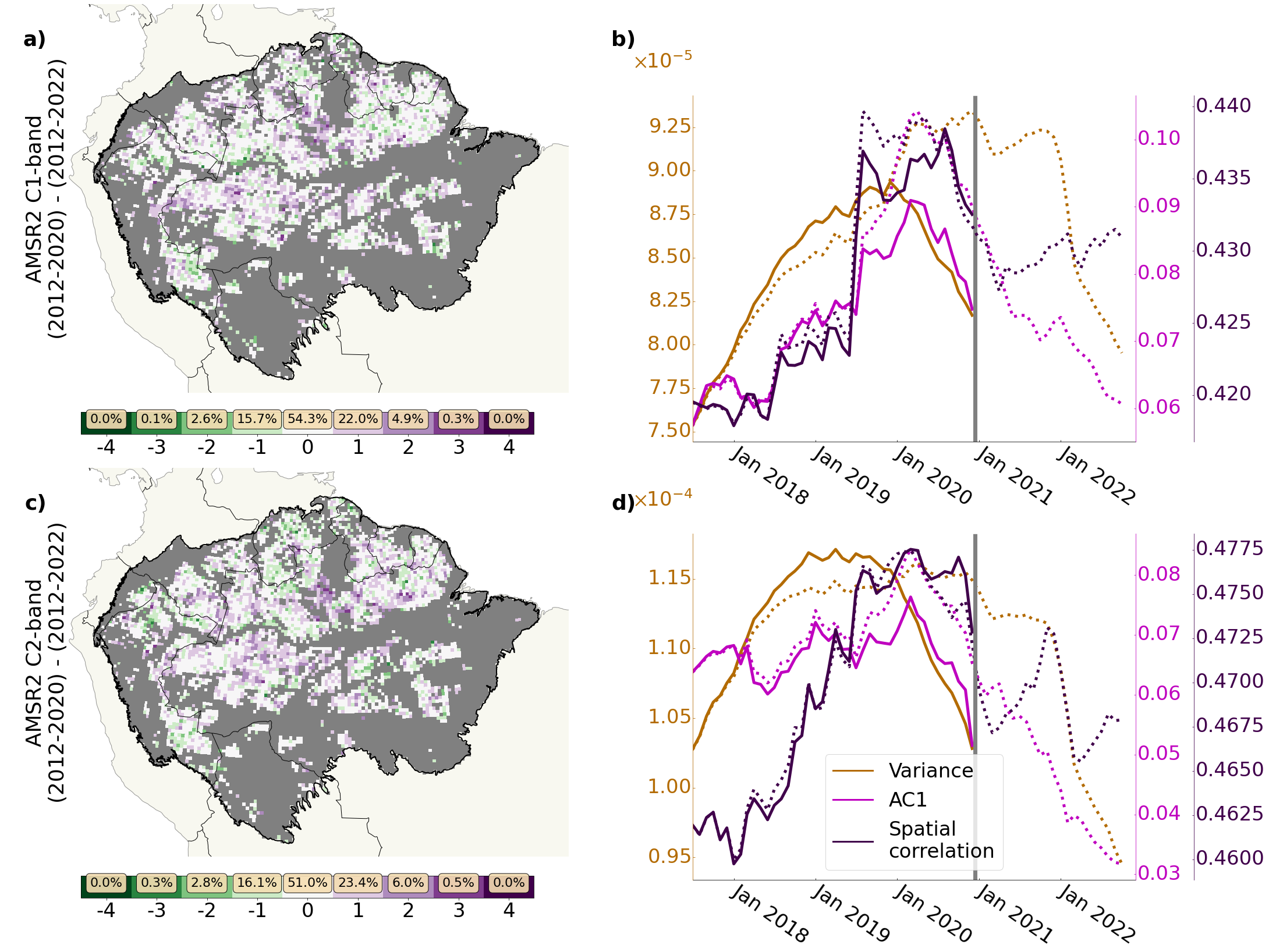}
	
	\caption{
		\textbf{Differences in indicators for AMSR2 induced by the period of analysis.}
		Since the data sets for excluding HLU are only available until 2020, the VOD data by AMSR2 was only analyzed until 2020 even though it is available until present times. 
		This Figure shows \textbf{a)} and \textbf{c)} the difference between the number of indicators with positive trends if analyzed until 2020 (as in the main text, Figure~\ref{fig:AMSRE_summary}) and \textbf{b)} and \textbf{d)} the average time series when analyzed until 2022 for AMSR2's band C and X, respectively. 
	}
	\label{fig:SI_AMSR2_2022}
\end{figure}

Since the data sets for excluding HLU are only available until 2020, the VOD data by AMSR2 was only analyzed until 2020 even though it is available until present times. 
The maps in Figure \ref{fig:SI_AMSR2_2022}\textbf{a)} and \textbf{c)} show the difference between the number of indicators with positive trends if analyzed until 2020 (as in the main text, Figure~\ref{fig:AMSRE_summary}) and until 2022 for AMSR2's band C and X, respectively. Differences mainly occur where sudden changes in VOD after 2020 can have a strong influence on the detrending method STL.
Since STL acts on sliding windows over the whole data set as is (to detrend) and over the separate months (for de-seasonalizing), the influence of strong changes after 2020 increase over time. 
This becomes apparent in the the spatially averaged time series in \textbf{b)} and \textbf{d)}, where all indicators start at the same level independent for both cases, where the analysis is performed until 2020 (solid lines) as well as until 2022 (dotted lines), but diverge afterwards.
Such increasingly strong discrepancies in the period until 2020, where the data is equivalent, must stem from STL, which over time becomes more sensitive to strong changes in later time steps due to its sliding window approach. Yet, this is only the case for strong changes in the VOD after 2020, as it could result from forest loss and Human interference. 
Hence, we argue that the exclusion of HLU based on the years until 2020 is not sufficient to analyze data thereafter.
This explanation is supported by high rates of forest loss since the end of 2020, with more than 18.000\,km$^2$ deforestation and 27.000\,km$^2$ degradation registered by the Instituto Nacional de Pesquisas Espaciais's monitoring programs \citep{inpe_terrabrasilis_2023}.
Hence, the results for AMSR2 analyzed until 2022 cannot be relied on (see Figure~\ref{fig:SI_AMSR2_2022}).

\FloatBarrier
\clearpage

\section{Investigation of resilience loss in AMSR2's X-band}

\subsection*{Text S7. The influence of forest loss for the years 2021 to 2022}

\begin{figure}
	\centering
	\includegraphics[width=\textwidth]{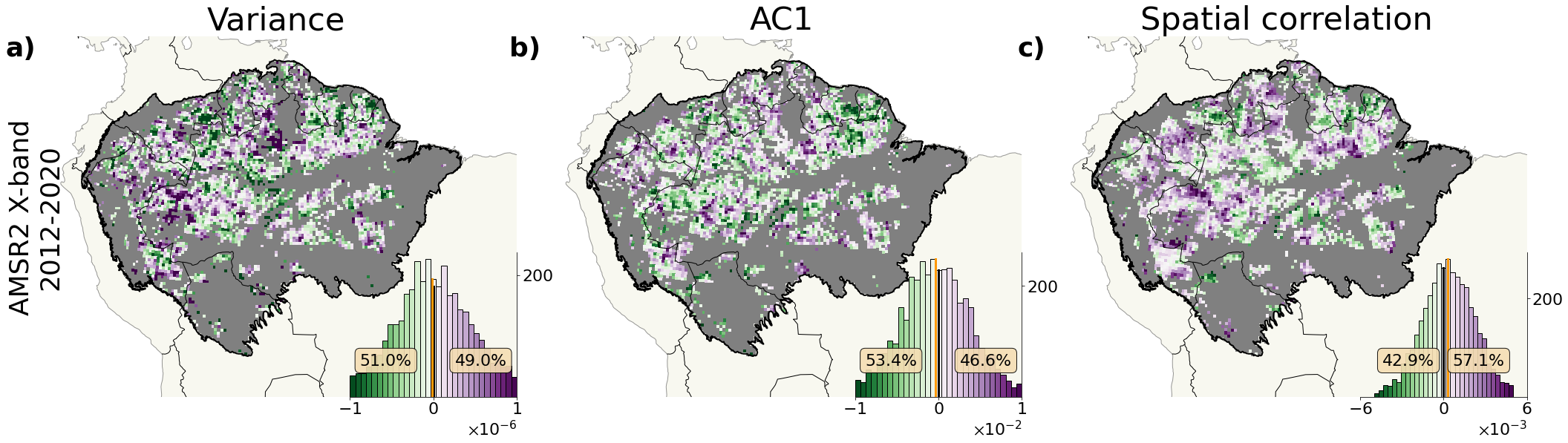}
	
	\caption{
		\textbf{Change in single indicators of CSD based on AMSR2's band X.}
		Trends in variance, AC1, and Spatial  Correlation for the X-band of AMSR2.
	}
	\label{fig:SI_AMSR2_bandX_1}
\end{figure}

\begin{figure}
	\centering
	\includegraphics[width=\textwidth]{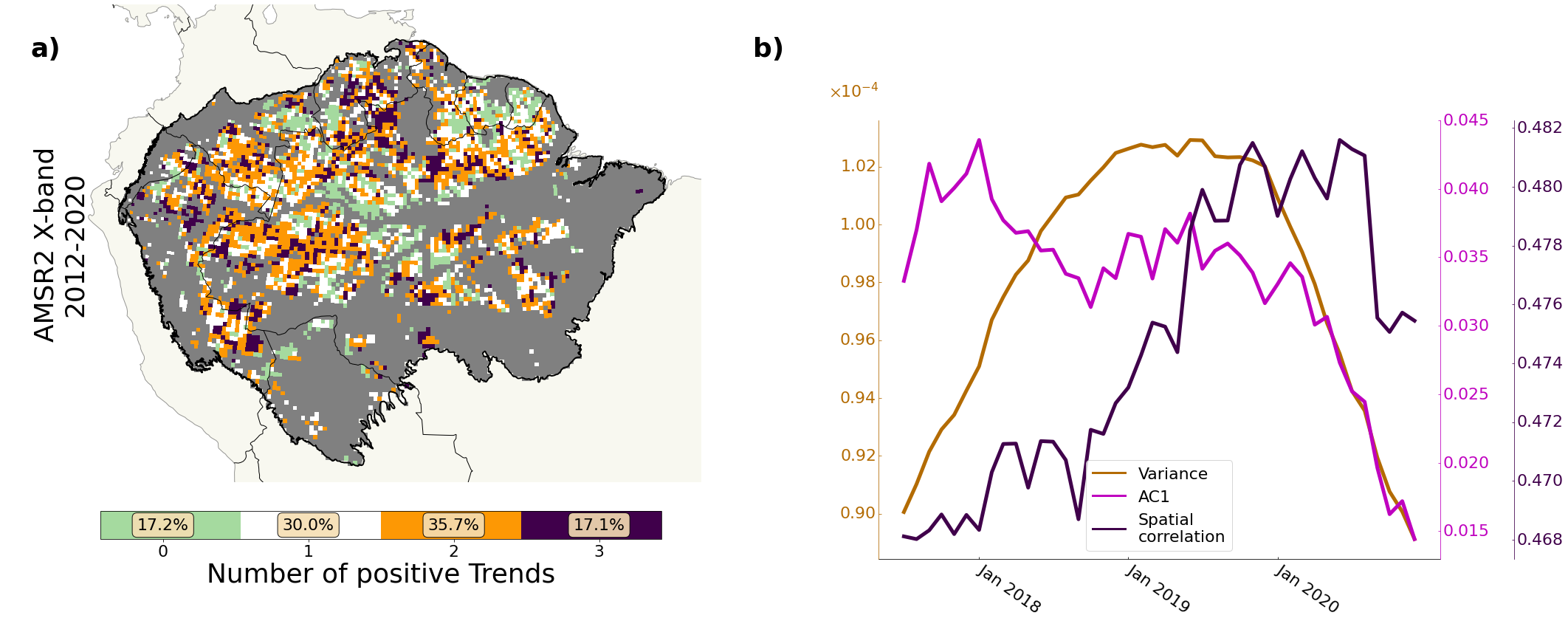}
	\caption{
		\textbf{Summary of indicators of CSD based on AMSR2's band X.}
		This Figure shows \textbf{a)} the number of indicators with a positive trend per cell and \textbf{b)} the spatially averaged evolution of the single indicators over time.
	}
	\label{fig:SI_AMSR2_bandX_2}
\end{figure}

The X-band of AMSR2 results from the observation of a shorter wavelength compared to the two C-bands. This leads to less sensitivity to aboveground biomass, which is why the results are less likely to indicate whether the ARF is loosing resilience or not. 
For completeness, Figures \ref{fig:SI_AMSR2_bandX_1} and \ref{fig:SI_AMSR2_bandX_2} show the results for the X-band.
While the spatial correlation is more often increasing than not, which is comparable to the results in the C-bands, this is not the case for variance and AC1 (Figure \ref{fig:SI_AMSR2_bandX_1}).
While the percentage of cells in Figure \ref{fig:SI_AMSR2_bandX_2}\textbf{a)} exhibiting a positive trend in all three indicators is still above what would be expected by random (12.5\,\%), it is far less than detected for AMSR2's C-bands.

\FloatBarrier
\clearpage

\section{Results proving robustness}		

\subsection*{Text S8. Robustness with respect to the parameterization of the STL algorithm}
\begin{figure}
	\includegraphics[width=\textwidth]{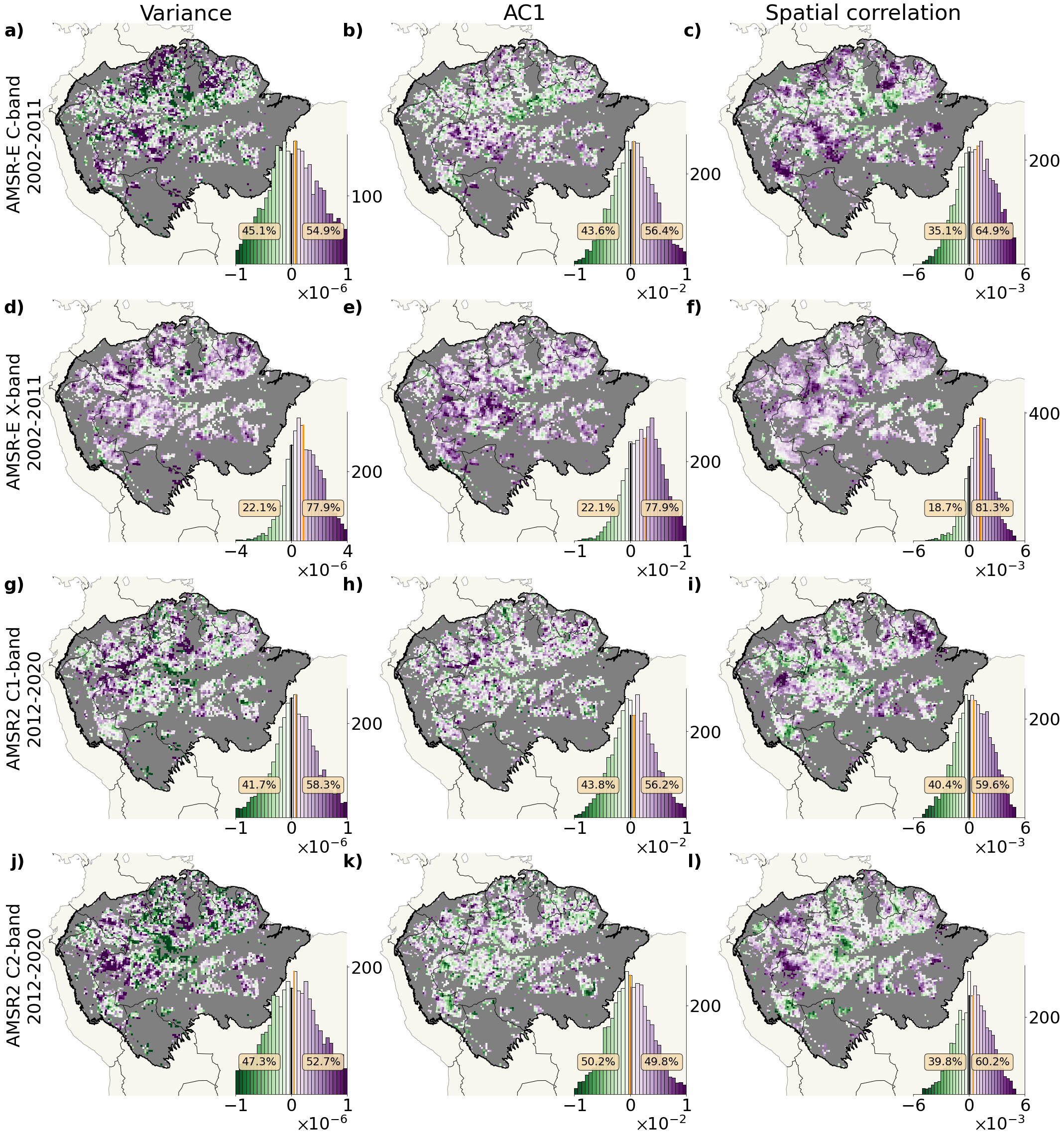}
	\caption{
		\textbf{Robustness with respect to parametrization of detrending and deseasonalization algorithm STL.}
		Trends in variance, AC1, and spatial correlation of the residual time series resulting from a different parametrization of the STL algorithm (seasonal smoother in STL set to 13 instead of 7).
	}
	\label{fig:SI_STL}
\end{figure}

\begin{figure}
	\includegraphics[width=\textwidth]{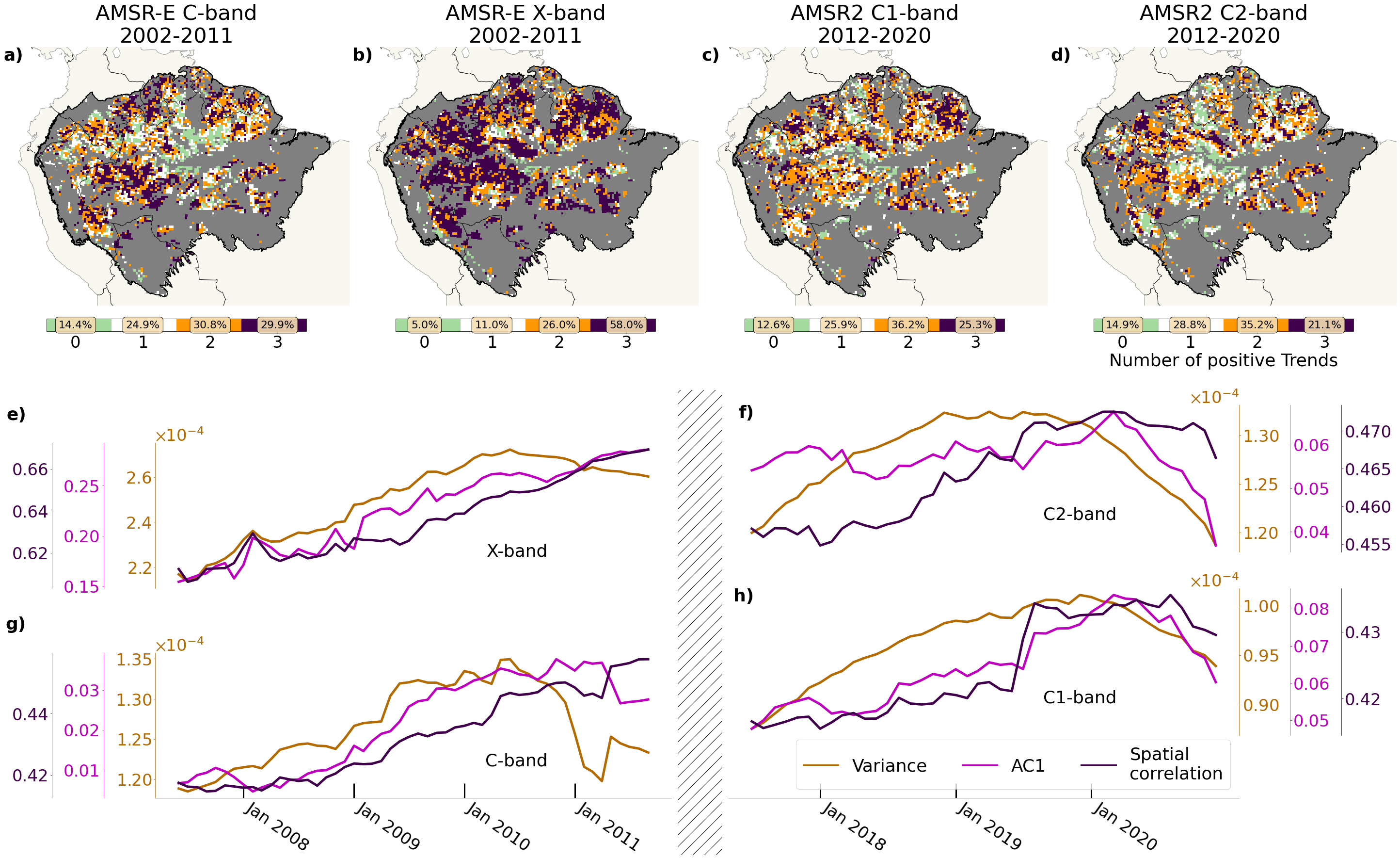}
	\caption{
		\textbf{Robustness with respect to parametrization of detrending and deseasonalization algorithm STL.}
		Summary of the resilience changes detected by the three indicators, here based on residual time series resulting from a different parametrization of the STL algorithm (seasonal smoother in STL set to 13 instead of 7).
	}
	\label{fig:SI_STL_summary}
\end{figure}

The results in Figures~\ref{fig:AMSRE_indicators} and \ref{fig:AMSRE_summary} are based on the residual found by the STL algorithm with default parameterization. Yet, Figure \ref{fig:SI_STL} and \ref{fig:SI_STL_summary} prove that the analysis is not influenced by this choice as the results are almost identical when the length of the seasonal smoother equals 13 (instead of 7).

\subsection*{Text S9. Robustness with respect to the size of the sliding windows}
\begin{figure}
	\includegraphics[width=\textwidth]{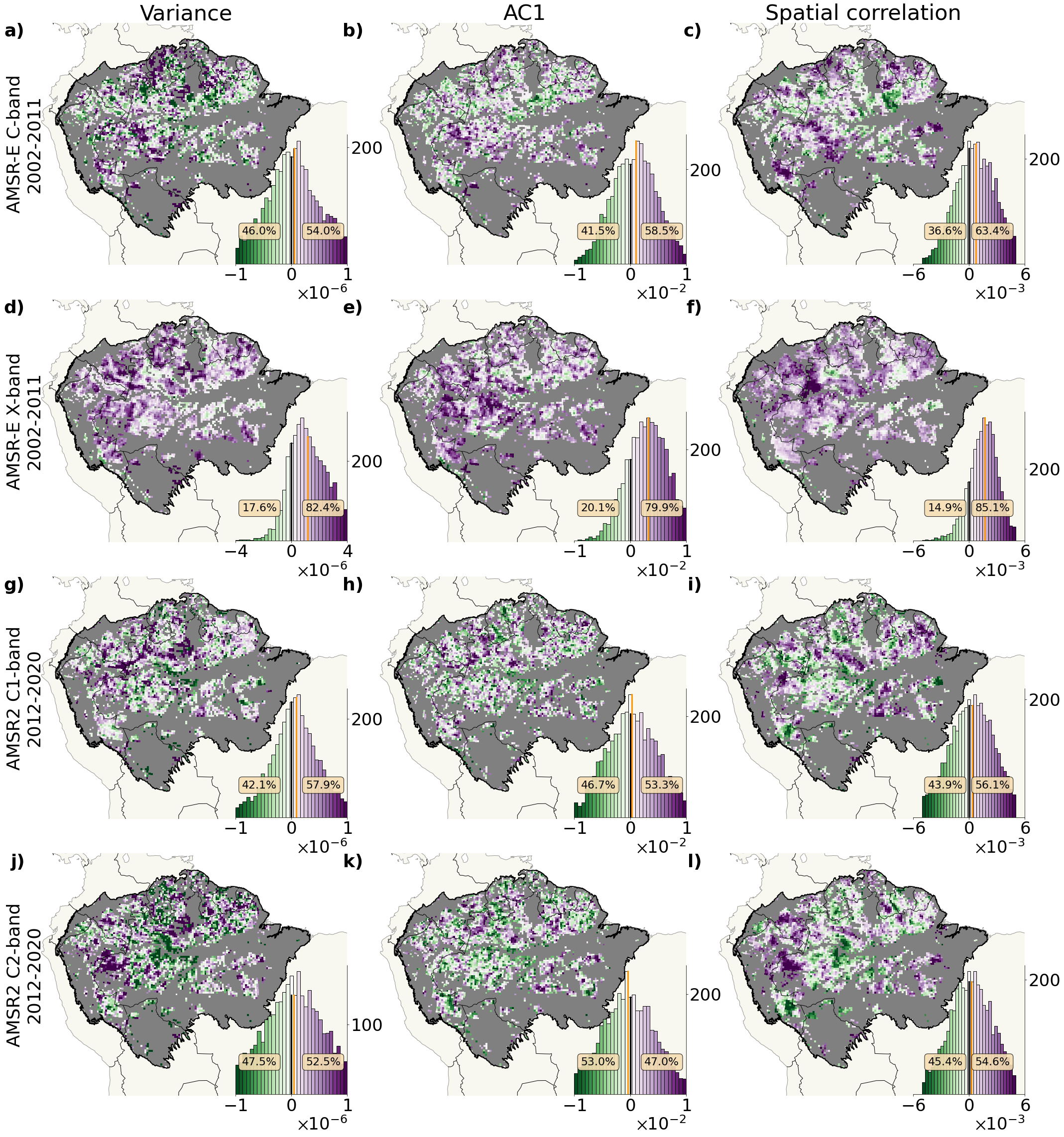}
	\caption{
		\textbf{Robustness with respect to the sliding window size for calculating the indicators of CSD.}
		Trends in variance, AC1, and spatial correlation based on sliding windows of 3 years.
	}
	\label{fig:SI_sws}
\end{figure}

\begin{figure}
	\includegraphics[width=\textwidth]{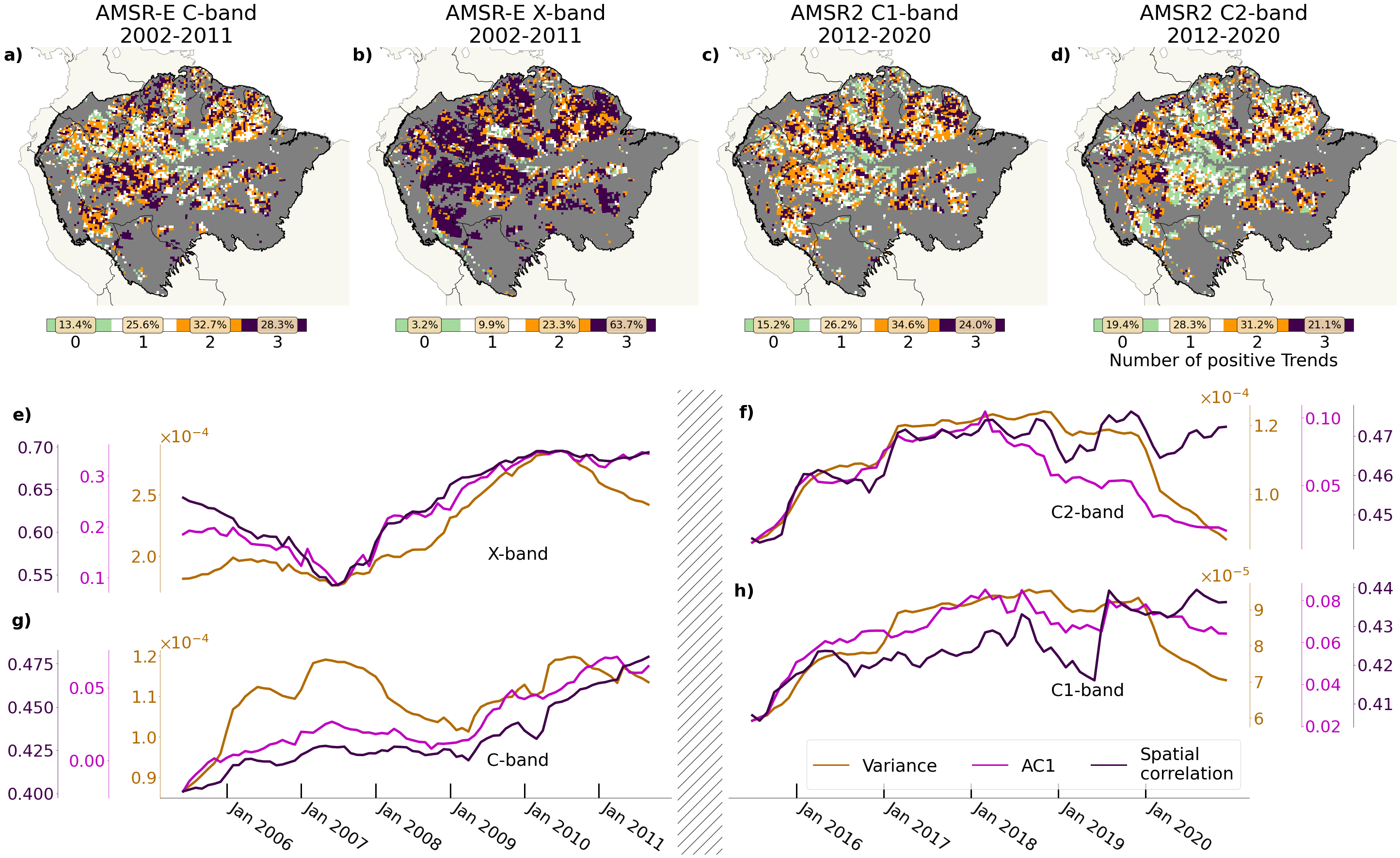}
	\caption{
		\textbf{Robustness with respect to the sliding window size for calculating the indicators of CSD.}
		Summary of the resilience changes detected by the three indicators, here  based on sliding windows of 3 years.
	}
	\label{fig:SI_sws_summary}
\end{figure}

The choice of the sliding window size is difficult for short time series as that of AMSR-E's and AMSR2's VOD. Yet, the results are robust w.r.t it, as Figure \ref{fig:SI_sws} based on sliding windows of 3 years is almost identical to Figure~\ref{fig:AMSRE_indicators} with windows of 5 years. This also holds for the summarizing Figure~\ref{fig:SI_sws_summary} when compared to Figure~\ref{fig:AMSRE_summary}.
It is to note that the spatially averaged time series of the three indicators are more noisy for the choice of smaller windows, which is due to the fact the the calculation of the indicators becomes more unstable when calculated on only 36 data points, corresponding to the 3 years. 
The right choice of the window size in an approach to resilience analysis like the one shown here is challenging, as the increase can only be determined reliably based on enough indicator time steps either. The fact that the results are robust with respect to the window size gives confidence in the choice of 5 years as default.

\subsection*{Text S10. Robustness with respect to the measure of change}
\begin{figure}
	\includegraphics[width=\textwidth]{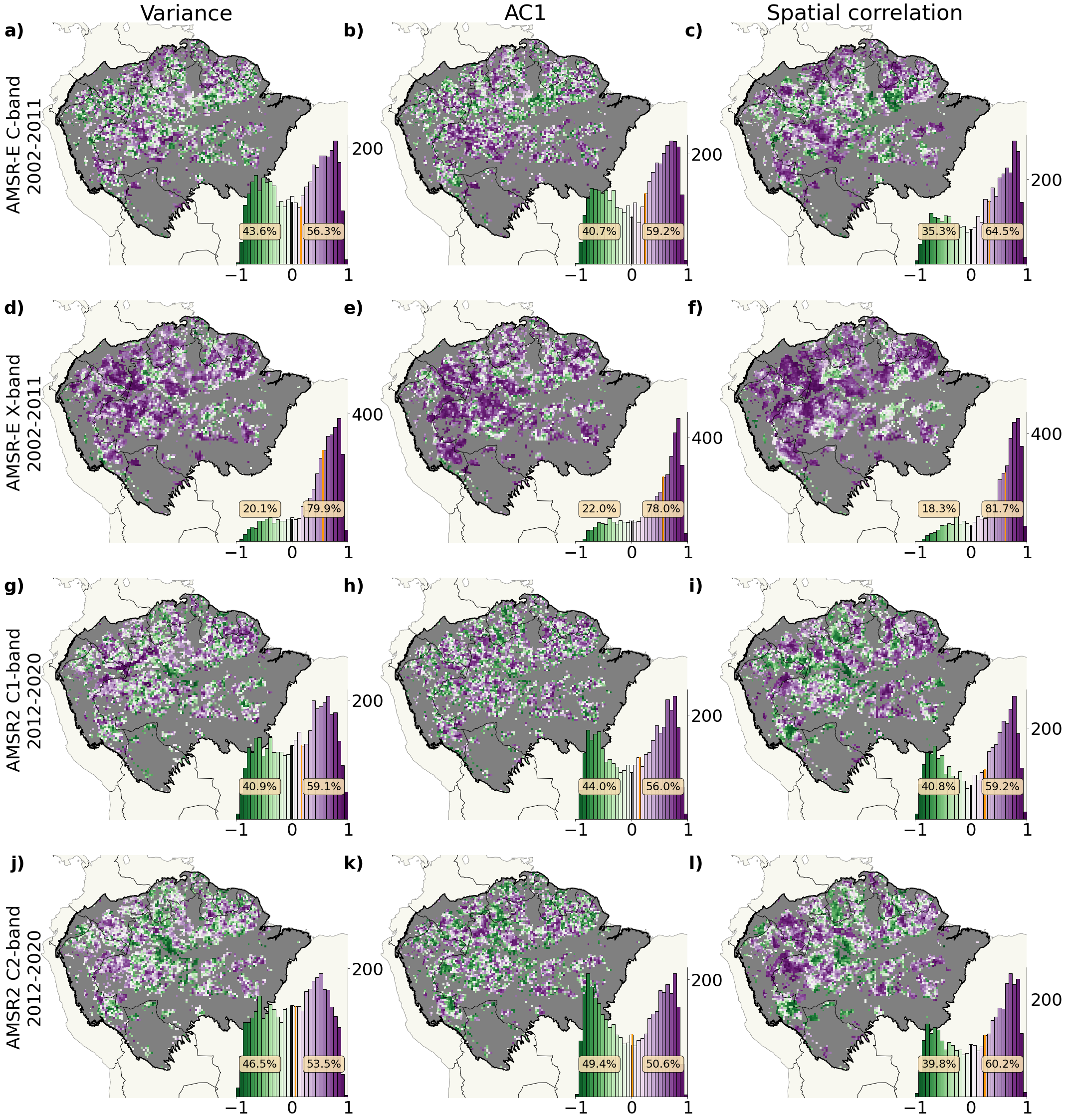}
	\caption{
		\textbf{Robustness with respect to the measure of increase.}
		The increase in the time series is quantified in terms of Kendall's $\tau$ of variance, AC1, and spatial correlation.
	}
	\label{fig:SI_KendallTau}
\end{figure}

In the main text, any change in indicators is measured in terms of linear trends. Yet, another option would be the Kendall's $\tau$ as a measure of steadiness of increase. Figure \ref{fig:SI_KendallTau} is the analogue of Figure~\ref{fig:AMSRE_indicators} but with increase in time series quantified in terms of Kendall's $\tau$.
In Figure~\ref{fig:AMSRE_summary}, increase in the indicators is quantified in terms of linear trends. The results are almost identical when measuring the indicators' increase by the Kendall $\tau$ rank coefficient. 

\begin{figure}
	\includegraphics[width=\textwidth]{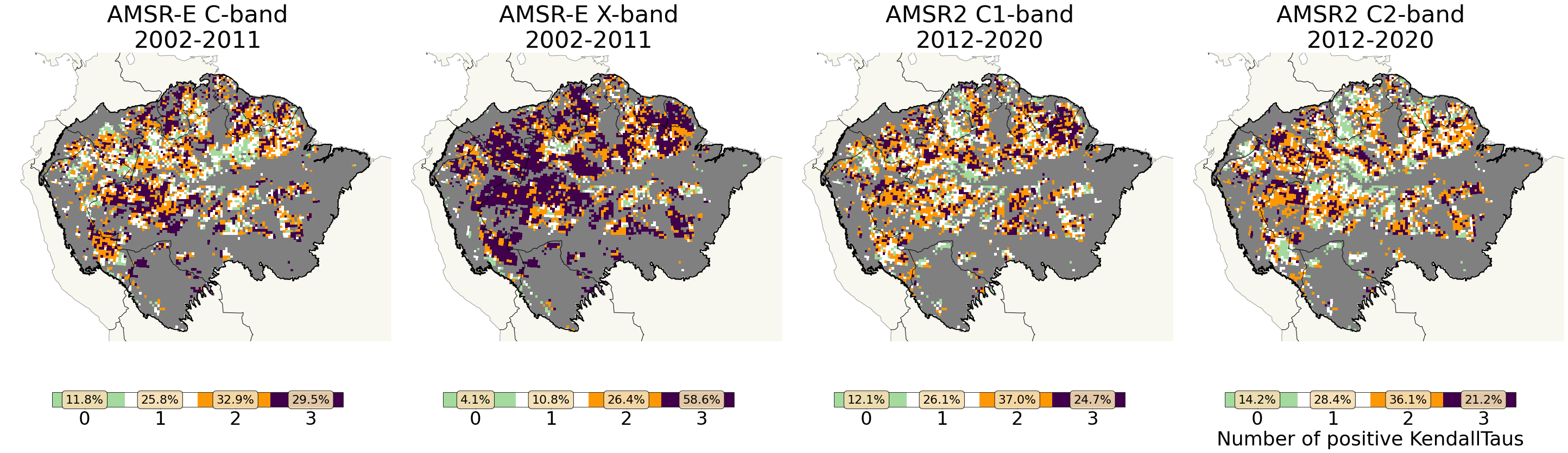}
	\caption{
		\textbf{Robustness with respect to the measure of increase.}
		Summary of the resilience changes detected by the three indicators, when the change in the time series is quantified in terms of Kendall's $\tau$ 
	}
	\label{fig:SI_KendallTau_summary}
\end{figure}

\subsection*{Text S11. Robustness with respect to the distance defining neighboring cells}
\begin{figure}
	\includegraphics[width=\textwidth]{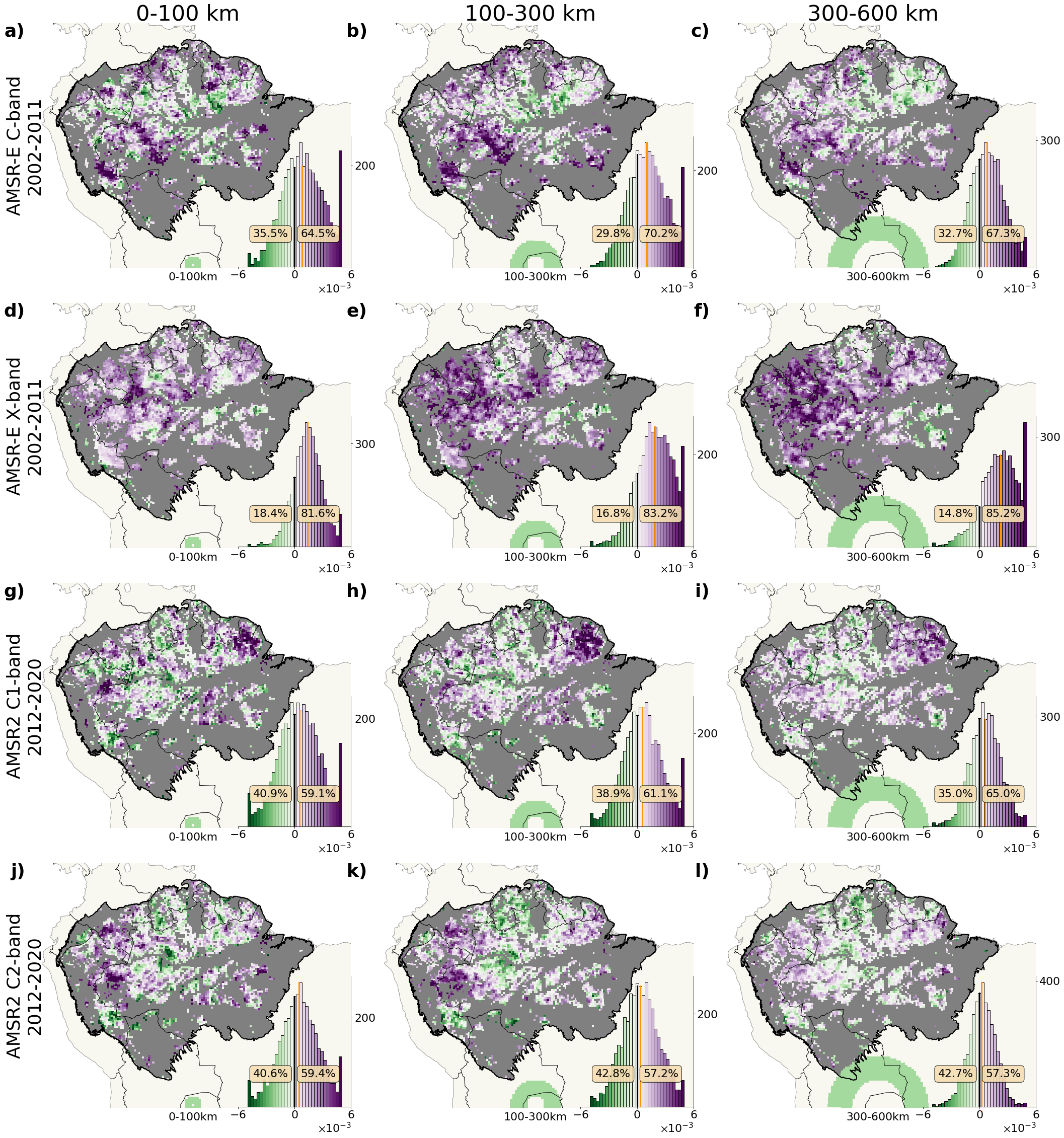}
	
	\caption{
		\textbf{Robustness of spatial correlation with respect to the definition of neighboring cells.}
		Trend of the spatial correlation for the different VIs comparing radii of 0-100, 100-300, and 300-600\,km as the range of cells taken into account as neighboring the center cell.
	}
	\label{fig:SI_distances}
	
\end{figure}

The definition of the spatial correlation depends on the distance taken into account when taking the average correlation with neighboring cells. Figure \ref{fig:SI_distances} shows the trends of the spatial correlation for the different VIs comparing different radii defining the set of neighboring cells. As expected, a greater number of neighbors spatially smooths the results, but does not influence the general spatial pattern of the trends of spatial correlation. The turquoise circle on the bottom of the map shows an exemplary circular band of neighboring cells.
In the main text, all results are based on a radius of 100\,km.

\FloatBarrier
\clearpage
\section{Potential driving forces and sources of bias}


\subsection*{Text S12. The influence of the South American Monsoon system as a potentially bistable system}
\begin{figure}
	\centering
	\includegraphics[width=.8\textwidth]{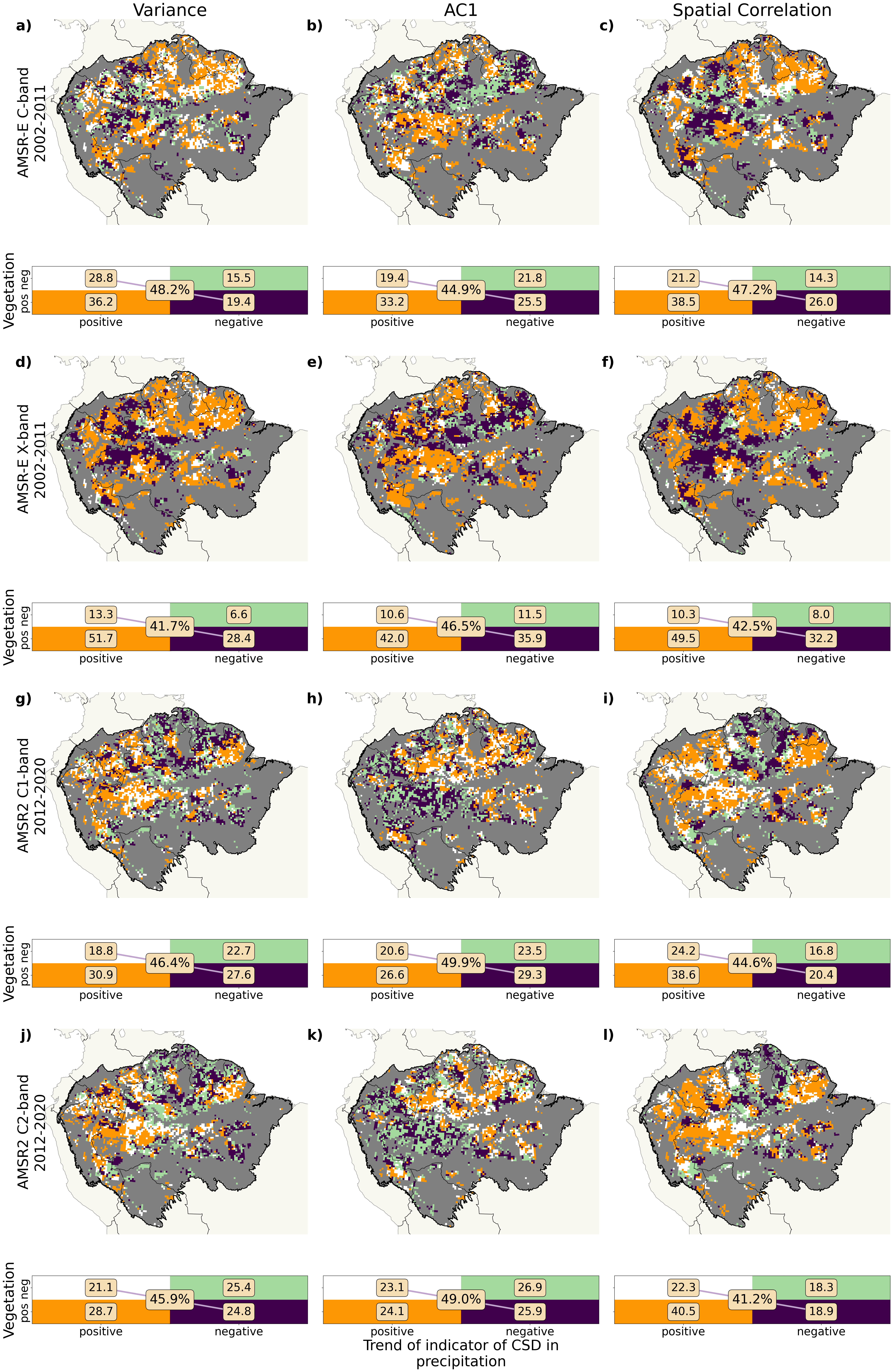}
	\caption{
		\textbf{The South American monsoon system as a potentially bi-stable system.}
		In this Figure, the signs of the trend of variance, AC1 and spatial are compared between VOD  and precipitation \citep{funk_quasi-global_2014}.
		The sum of the diagonal cells in boxes is the amount of cells where the values \textit{differ} in sign for the VIs and precipitation.}
	\label{fig:SI_VegPrecip_CSD_indicators}
\end{figure}

In the Amazon basin, the South American monsoon system and the atmospheric moisture recycling form a complex and potentially bi-stable system. Hence the monthly precipitation itself could exhibit signs of CSD, which would then drive vegetation changes that are wrongly interpreted as the destabilization of the vegetation. 
In Figure \ref{fig:SI_VegPrecip_CSD_indicators}, the signs of the trend of variance, AC1 and spatial are compared between VOD and precipitation \citep{funk_quasi-global_2014}.
We argue that due to the high amount of cells where the values differ in sign for the VIs and precipitation (sum of diagonal cells in boxes expected to be 50\,\% for random assignment), the observed changes in the indicators are likely a measure of critical slowing down in the vegetation and not only a direct effect of variability in precipitation.		

\subsection*{Text S13. The influence of changes in precipitation as a potential driving force}
\begin{figure}
	\centering
	\includegraphics[width=.8\textwidth]{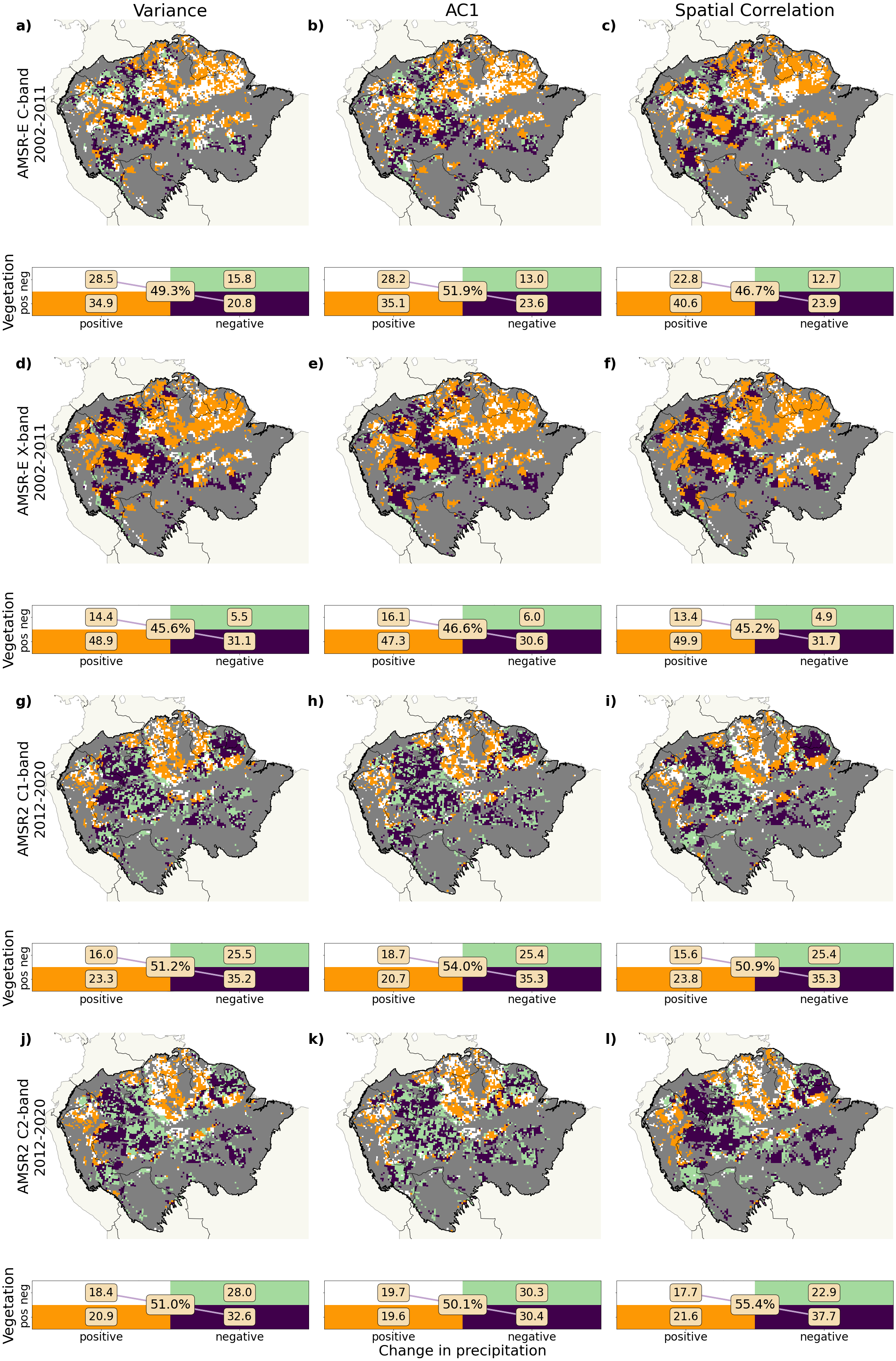}
	\caption{
		\textbf{Precipitation as the main potential driver of resilience loss in the ARF.}
		In this Figure, the signs of the trend of variance, AC1 and spatial are compared between VOD and the direct trends of precipitation.
		The sum of the diagonal cells in boxes is the amount of cells where the values \textit{agree} in sign for the VIs and precipitation.
	}
	\label{fig:SI_precipitation_attribution}
\end{figure}

If precipitation is the main driver of ARF resilience loss, one would expect that in regions with stronger decreases in precipitation (negative trend), more critical slowing down can be observed.
This comparison in displayed in Figure \ref{fig:SI_precipitation_attribution}.
In parts, decreased precipitation could potentially explain the destabilization detected in the western Amazon basin by both bands of AMSR-E as well as in the western and northeastern regions detected by AMSR2's bands.
However it is important to note here that the number of cells with an increasing amount of precipitation and decreasing indicators and those with decreasing amount of precipitation and increasing indicators add up to only about 50\,\%, which is what is to be expected for a random assignment of positive and negative trends in both.

\subsection*{Text S14. The influence of changes in temperature}
\begin{figure}
	\centering
	\includegraphics[width=.8\textwidth]{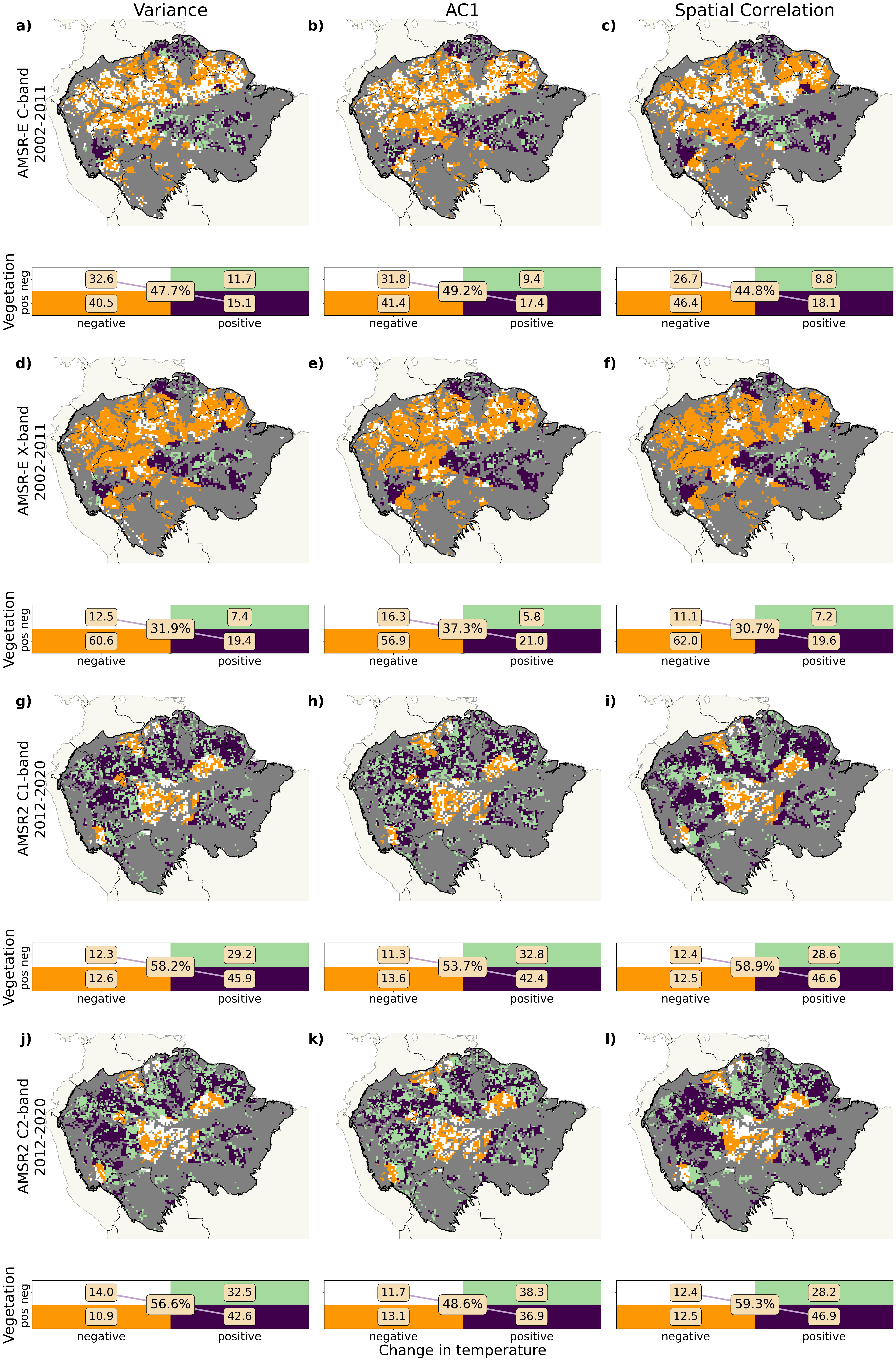}
	\caption{
		\textbf{Increasing temperature as a potential driver.}
		In this Figure, the signs of the trend of variance, AC1 and spatial are compared between VOD and the direct trends of temperature.
		The sum of the diagonal cells in boxes is the amount of cells where the values \textit{agree} in sign for the VIs and precipitation.
	}
	\label{fig:SI_temperature_attribution}
\end{figure}

Instead of decreasing precipitation (see Figure~\ref{fig:SI_precipitation_attribution}, an increase in temperature can also lead to vegetation destabilization through water stress by higher evapo-transpirative demand.
If this was the driver of the CSD, increasing temperatures must coincide with positive trends in the indicators of CSD in Figure \ref{fig:SI_temperature_attribution}.
Yet changes in temperature cannot explain the detected destabilization in the years 2002 to 2011.
For the time period of AMSR2, destabilization detected in the northeast (band C1) and the east (band C2) might at least partially be accounted for rising temperatures (the percentage of cells with rising/declining temperatures and in-/decreasing indicators falls between 48.6\,\% and up to 59.3\,\%).

\subsection*{Text S15. The potential bias induced by saturation}
\begin{figure}
	\centering
	\includegraphics[width=.8\textwidth]{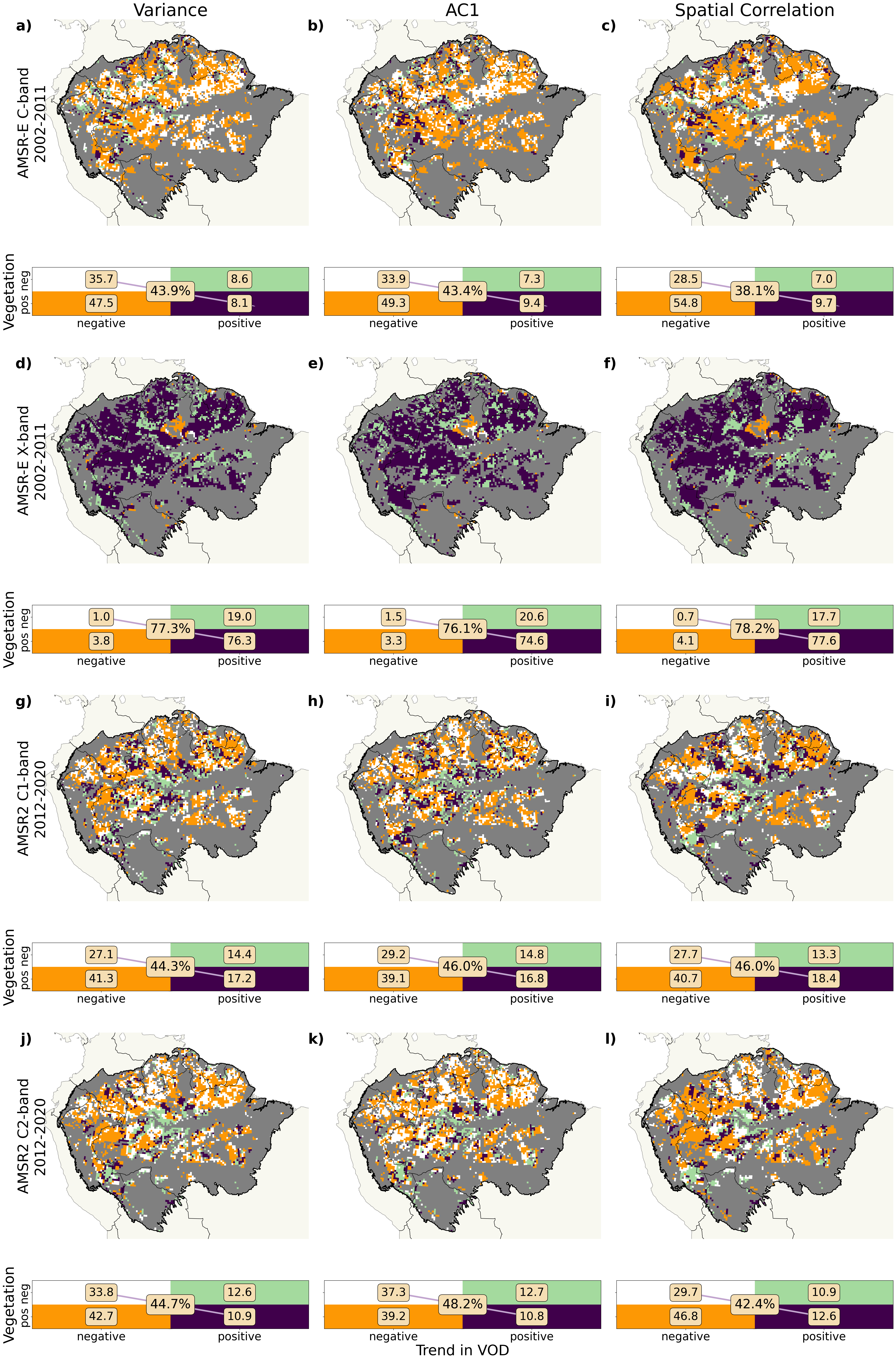}
	\caption{
		\textbf{Saturation as a source of artifacts.}
		In this Figure, the signs of the trend of variance, AC1 and spatial are compared between VOD and the direct trends of the VIs.
		The sum of the diagonal cells in boxes is the amount of cells where the values \textit{agree} in sign for the VIs' CSD indicators and the VIs themselves.
	}
	\label{fig:maxDist100_SI_artifact_saturation_attribution}
\end{figure}

In case of optical sensors, their saturation over high biomass regions often renders the VIs insensitive to the subtle changes due to perturbations that a resilience analysis based on CSD relies on.
The argument is that with decreasing biomass, the VIs become more sensitive, which then translates into a signal of CSD as a direct artifact of the higher perception of perturbations.
Even though VOD results from observations of microwaves, we want to address the question of saturation in VOD. 
As Figure \ref{fig:maxDist100_SI_artifact_saturation_attribution} shows, we cannot exclude this as an artifact for AMSR-E's band C as well as AMSR2's two bands, as they hardly observed any cells with increasing biomass.
Interestingly, AMSR-E's band X detected an increase in biomass as well as a loss of resilience in most of the cells (77.3\,\% to 78.2\,\%). Thus, its signal of destabilization of almost the entire ARF during the years 2002 to 2011 is not an artifact of saturation of the band.

\FloatBarrier
\clearpage

\end{document}